\documentclass[reprint,amsmath,amssymb,aps,pre]{revtex4-1}

\usepackage{graphicx}% Include figure files
\usepackage{dcolumn}% Align table columns on decimal point
\usepackage{bm}% bold math
\usepackage{hyperref}% add hypertext capabilities

\graphicspath{{figures/}}

\begin{document}
\title{Numerical Estimation of Limiting Large-Deviation Rate Functions}
	\author{Peter Werner}
	\email{peter.werner@uni-oldenburg.de}
	\author{Alexander K. Hartmann}%
	\email{a.hartmann@uni-oldenburg.de}
	\affiliation{
		Institut f\"ur Physik, Universit\"at Oldenburg, 26111 Oldenburg, Germany
	}
\date{\today}
\begin{abstract}
  For statistics of rare events in systems obeying a
  large-deviation principle, the rate function is a key quantity.
  When numerically estimating the rate function one is always
  restricted to finite system sizes. Thus, if the interest is in the
  limiting rate function for infinite system sizes, first, several
  system sizes have to be studied numerically. Here rare-event algorithms
  using biased ensembles give access to the low-probability region.
  Second, some kind of system-size extrapolation has to be performed.
 
  Here we demonstrate how rare-event importance sampling schemes
  can be combined with multi-histogram reweighting, which allows for
  rather general applicability of the approach, independent of specific
  sampling algorithms.
  We study two ways of performing the system-size extrapolation, either directly
  acting on the empirical rate functions, or on the 
  scaled cumulant generating functions, to obtain the infinite-size limit.
  The presented method is demonstrated for a binomial distributed variable
  and the largest connected component in Erd\"os-R\'enyi random graphs. 
  Analytical solutions are available in both cases for direct comparison. 
  It is observed in particular that phase transitions appearing in the biased 
  ensembles can lead
  to systematic deviations from the true result.
\end{abstract}
	
	\maketitle
	
\section{Introduction}
%-Background & General motivation
Let $S$ be an extensive random variable of a system at scale 
$\mathcal{N}$ that is under consideration.
The scale $\mathcal{N}$ could be the duration of a process or the 
number of individual system components.
For a very simple example, let $S$ be the number of rainy 
days in a region within	a total time span of $\mathcal{N}$ days.
A corresponding intensive quantity is then $s=S / \mathcal{N}$, 
i.e. the fraction of rainy days. Since the number of days varies
with time, a stochastic modeling is natural, described by a
probability distribution $p(s; \mathcal{N})$.
Often such distributions follow the \emph{large-deviation principle}
\cite{denHollander2000,Touchette_2009,Touchette_2012},
which means that there exists a so-called rate function $I(s)$ such that
\begin{equation}\label{eq:large:deviation:principle}
  p(s; \mathcal{N}) \asymp e^{-\mathcal{N} I(s)}\,.
\end{equation}
This is equivalent to asking whether the limit
\begin{equation}\label{eq:definition:rate:function}
  \lim_\mathcal{S\rightarrow\infty} -\frac{1}{\mathcal{N}} \ln p(s; 
  \mathcal{N}) = I(s)
\end{equation}
exists?
There is an analytical solution for a simple model of the rainy day
example, as will be 
discussed below, but for more complex systems numerical methods are a 
viable alternative. Here many different systems have been investigated
in the physics community like
random graphs \cite{Engel_2004,Hartmann_2011,Hartmann_2017},
exclusion processes \cite{Giardina_2006, Lecomte_2007},
the contact process \cite{Hidalgo_2017}, spin glasses
\cite{pe_sk2006}, Brownian \cite{Cerou_2007, Agranov_2020} or
fractional Brownian motion \cite{fBm_MC2013,fBm_PA2024, Smith_2022},
traffic models \cite{nagel_schreckenberg2019}, the Kardar-Parisi-Zhang
equation \cite{kpz2018,kpz_HMS2019,kpz_long2020},
chaotic maps \cite{Smith_chaotic_maps_2022},
non-equilibrium work processes \cite{Hartmann_2014,Werner_2021,Werner_2024},
and many more. Here one has to
take into account that the probability density $p(s; \mathcal{N})$
following Eq.~(\ref{eq:large:deviation:principle}) becomes exponentially
small in $\mathcal{N}$ almost everywhere. Therefore,
in order to obtain a good estimate of the rate
function $I(s)$, rare-event samples have to be obtained
from the distribution tails, which is a numerical challenge.
	% Overview Numerical Methods
For this purpose, various large-deviation
algorithms  and methods \cite{bucklew2004,bouchet2019} have been 
devised.
One class of approaches are based on population dynamics, referred
to as \emph{cloning} or \emph{splitting}
\cite{Giardina_2006, Lecomte_2007, Tailleur_2009, Hidalgo_2017},
with Adaptive Multilevel Splitting (AMS)
\cite{Cerou_2007, Brehier_2015, Cerou_2019} as a notable variant.
In these, a set of systems is evolved simultaneously and at certain points 
systems are discarded and/or duplicated based on an importance function 
that biases the dynamics towards rare-events.
Another approach, the adaptive power method \cite{Coghi_2023}, is centered 
around estimating the eigenvalue of the systems tilted transition matrix, 
which allows to get the rate function by means of the G\"arnter-Ellis 
theorem \cite{Touchette_2009}.
In some methods, the maximum likelihood pathway of a rare-event, the so 
called \emph{instanton}, is numerically determined by mapping to a 
classical system, to gain insight into a systems large-deviation properties 
\cite{HartmannC_2019, Grafke_2019, Alqahtani_2021}.
The final class of approaches are importance-sampling methods 
\cite{Glasserman_1997, Hartmann_2014, Hartmann_2011, Hartmann_2017, Werner_2021, Werner_2024}.
Here, data points are sampled according to some Monte-Carlo scheme but with
respect to a tilted or biased distribution to obtain improved estimates 
in the original distributions tails.
This later class of algorithms are of relevance here.

	%Purpose of this paper
The purpose of this paper is to demonstrate how multi-histogram
reweighting \cite{Ferrenberg_1988, Ferrenberg_1989, kumar1992,
  Ferrenberg_1995, bereau2009}
can be effectively  combined with importance-sampling large-deviation
techniques and finite-size fitting
extrapolation \cite{Hidalgo_2017, Nemoto_2017,
  Hidalgo_2018} to obtain a rate function estimate.  

%-Paper structure
The paper is structured as follows: We start by explaining methods for
estimating rate functions.  Subsequently, two
case studies are presented while applying
the above mentioned approaches. First, a simple binomial
distributed case is considered. Second, we investigate
the more complex case of the largest
connected component in Erd\"os R\'enyi (ER) random graphs.
For these investigations, the general behavior of the
approaches in relation to their  parameters are demonstrated.
We conclude with an overarching discussion.
	
\section{Rate Function Estimation Methods}
\subsection{Empirical Rate Function at a Large Scale}
A naive way to estimate a rate function is to perform large-deviation
simulations at a sufficiently large system scale $\mathcal{N}$, ideally 
chosen such that the empirical determined finite-size corrections are 
insignificant compared to the desired magnitude of statistical errors.

An estimate of the underlying probability distribution $p(s, \mathcal{N})$ 
for the quantity of interest in form of a histogram $h(s;\mathcal{N})$
is straightforward to calculate, see \cite{Hartmann_2014} for an example.
From this, the empirical rate function is given by:
\begin{equation}\label{eq:empirical:rate:function}
  I(s;\mathcal{N})=-\frac{\ln(h(s; \mathcal{N}))}{\mathcal{N}}.
\end{equation}
This approach assumes that the chosen system size is ``large enough''
which can roughly be
tested by performing at least some simulations for a larger size
like $2\mathcal{N}$
and verifying that the results do not change as compared to the statistical
error.

A minor but very common
drawback of this method is that the histogram requires to
select an appropriate binning, which can limit the spacial resolution in
regions with less data samples.
Statistical error propagation, on the other hand, is relatively easy
accomplished with this method via simple Gaussian error-propagation.
The approach presented below is histogram-free and involve
error propagation, too.
	
\subsection{Direct Extrapolation of the Empirical Rate Function}
An extension of the previous method is to obtain data for various
system scales $\mathcal{N}$
and subsequently calculate the empirical rate function
in Eq. \eqref{eq:empirical:rate:function} at each of them.
Assuming that the finite-size effects behave like a power-law
as a function of $\mathcal{N}$, i.e., 
\begin{align}\label{eq:power:law:s:space}
  I(s; \mathcal{N}) =& \tilde{I}(s) + 
  a(s) \mathcal{N}^{-b(s)}\\
  &\text{with}\; b(s) \ge 0, \nonumber
\end{align}
allows a fit for selected values of $s$ over the system scales $\mathcal{N}$, 
where $\tilde{I}$, $a$ and $b$ are $s$-dependent  fit parameters.
Thus, for every considered value of $s$ an independent fit is performed.
The extrapolated rate function is $\tilde{I}(s)$.
There is no a-priory argument, why a power law is an appropriate choice in 
all cases. 
However, such shape of finite-size dependencies is widespread in statistical
physics and
it has observed already  in large-deviation 
contexts \cite{Hidalgo_2017, Nemoto_2017, Edelman_2016}.
The discussion following Eq. \eqref{eq:power:law} below will elaborate this 
briefly.
When the measured quantity $S$ only takes values from a finite 
set of elements that scales with $\mathcal{N}$, this approach
comes with a further drawback.
Since the empirical rate function is then evaluated at discrete points 
$s=S/\mathcal{N}$, too, which must be common at all system scales 
$\mathcal{N}$, the spatial resolution of the rate function estimate will be 
limited by the resolution at smallest system scale.
	
\subsection{SCGF Transformation with Multi-Histogram Reweighting}
Instead of using Eq. \eqref{eq:power:law:s:space} for a direct 
estimate of the rate function, the approach presented here is 
centered around the approximation of the \emph{scaled cumulant 
  generating function} (SCGF) that is given by
\begin{equation}\label{eq:scgf}
  \varPsi(q) := \lim_{\mathcal{N}\rightarrow\infty} 
  \frac{1}{\mathcal{N}} \ln \left\langle e^{q \mathcal{N} s} 
  \right\rangle.
\end{equation}
From this estimate, the rate function is obtained via the 
Gärtner-Ellis theorem \cite{Touchette_2009, Touchette_2012},
which states that the rate function is the 
Legendre-Fenchel (LF) transformation of the SCGF:
\begin{equation}\label{eq:legendre:fenchel:transformation}
  I(s) = \sup_q \left[ qs - \varPsi(q)\right].
\end{equation}
Note that the application of this transformation assumes that the
underlying distribution follows the large-deviation principle
Eq.~(\ref{eq:large:deviation:principle}).
Furthermore, it
will only yield the convex envelope
of the rate function if the SCGF is non-convex \cite{Touchette_2009}.
There are benefits from taking this detour over the SCGF, compared to the 
direct methods.
First, no histogram binning is required to calculate the expectation value
in Eq. \eqref{eq:scgf}, which can just be approximated by an average of the 
data.
Second, should the distributions undergo substantial changes when 
increasing the system scale $\mathcal{N}$, i.e., strong finite-size effects,
Eq. \eqref{eq:scgf} is easier to extrapolate by means of a fit than the 
empirical rate function in Eq. \eqref{eq:empirical:rate:function} (see 
below).

The starting point is to perform numerical simulations
of the system of interest, for several
scales $\mathcal{N}$ and various
suitably chosen values of the \emph{bias parameter} $q$,
which appears already in
the definition (\ref{eq:scgf}) of the SCGF. The simulations generate
\emph{exponentially-biased} 
data series $(S^{q}_1,\ldots, S^{q}_{N_q})$ of $N_q$ values of
the system quantity $S$ of interest.
For convenience, the data points are assumed to be uncorrelated.
Note that for simplicity of notation
we do not denote the system-scale dependence
of the data series. Exponentially biased means here that for the given value
$q$ of the bias parameter
 the data points are
 obtained according the biased probability
 \begin{equation} \label{eq:prob:biased}
   \tilde{p}(S; \mathcal{N}, q)= \frac{p(S;\mathcal{N}) \exp(q S)}
   {\mathcal{Z}(q,\mathcal{N})}\,,
 \end{equation}
 instead of the original distribution $p(S;\mathcal{N})$.
Note that $\tilde{p}(S;\mathcal{N}, q=0)=p(S;\mathcal{N})$ for 
the unbiased case.
 Such biased data can be generated using corresponding large-deviation 
algorithms \cite{bolhuis2002,align2002,bucklew2004,Hartmann_2014}.
The exponential bias $\exp(q S)$ is chosen to explicitly match 
the functional form of the expectation value argument in Eq. 
\eqref{eq:scgf} and is frequently used in other previous studies that 
employ large-deviation algorithms \cite{Hartmann_2014, Werner_2021}.
In these studies a temperature like 
parameter $\theta$ is usually denoted in the exponential bias, 
which relates to the parameter $q$ through $q = - 1/\theta$.
In an importance sampling context, this setup is also referred to as an 
exponential change of measure \cite{Touchette_2012}.
Since the present
work is concerned with the data analysis and extrapolation rather
than the actual simulations, we refer to the literature on large-deviation
simulation algorithms.

The next step is to estimate from the numerical results
the SCGF at finite size $\mathcal{N}$, namely $\varPsi(q; \mathcal{N})$, and
the \emph{tilted} expectation
value $\mu(q; \mathcal{N}):= \varPsi'(q; \mathcal{N})$, i.e.,
the derivative of $\varPsi$ with respect to $q$. The
 latter one is needed to conveniently perform the LF
transform, see below. We start with $\mu(q; \mathcal{N})$ and obtain

\begin{align}
  \mu(q; \mathcal{N}) &:= \frac{\partial}{\partial q}
  \varPsi(q; \mathcal{N}) \label{eq:moments:scgf:derivative}\\
  &=\frac{\left\langle S e^{q S} 
    \right\rangle_{p(S;\mathcal{N})}}{\mathcal{N}\left\langle e^{q 
      S}\right\rangle_{p(S; \mathcal{N})}} \nonumber \\
  &= \frac{\left\langle S \right\rangle_{\tilde{p}(S; \mathcal{N}, 
      q)}}{\mathcal{N}} \label{eq:biased:moments}\\
  &= \left\langle s \right\rangle_{\tilde{p}(S; \mathcal{N}, q)} 
  \nonumber,
\end{align}
at finite system scales $\mathcal{N}$.
Details of the calculations for Eq. \eqref{eq:biased:moments} are shown 
in appendix \ref{ap:biased:moments}.
The expectation values
$\left\langle \dots \right\rangle_{\tilde{p}(S; \mathcal{N}, q)} $ are
with respect to the biased distribution in Eq. (\ref{eq:prob:biased})
 with the bias parameter $q$.

Thus, $\mu(q; \mathcal{N})$ can be estimated for a given size $\mathcal{N}$
directly for any value of $q$ where simulations have been performed
But large-deviation simulations can be computationally costly, therefore 
it is impractical to perform these for many values of $q$ at 
various scales $\mathcal{N}$. If the number of values $q$ is large,
this would be tedious.
To address this issue, we apply histogram reweighting, also often 
referred to as Ferrenberg-Swendsen reweighting
\cite{Ferrenberg_1988, Ferrenberg_1989, kumar1992,Ferrenberg_1995}, to obtain
estimates at arbitrary values of $q$.
Actually, in the form that is presented
below, the approach is histogram-free, which is very flexible.
To ensure that the reweighting procedure works correctly, the requirement 
arises that the spacing between bias values $q$ during the simulations 
is sufficiently small and therefore the relevant distribution support 
is decently covered by the sampled data.
However, this depends severely on the investigated model and has to be 
determined empirically in each case.

Let $q_i$ with $i = 1, \dots, K$ be the bias values used
within the simulations, i.e., $N_{q_i}$ denotes the number of
sampled data points for the $i$'th simulation.
An expectation value of a quantity $\mathcal{O}=\mathcal{O}(S)$,
i.e., an arbitrary  function depending on $S$, obtained 
at arbitrary bias $q$  is then given by \cite{kumar1992}
\begin{equation}\label{eq:fs:expectation:value}
  \left\langle \mathcal{O}\right\rangle_{\tilde{p}(S; 
    \mathcal{N},q)}= \sum_{i=1}^{K} \sum_{a=1}^{N_{q_i}}
  \frac{\mathcal{O}_a^{q_i} e^{q S_a^{q_i} +f_q}}{\sum_{k=1}^{K} 
    N_{q_k} e^{q_k S_a^{q_i} + f_{q_k}}}\;,
\end{equation}
where $\mathcal{O}_j^{q_i}=\mathcal{O}(S_j^{q_i})$.
The normalization constant $f_q$ is determined by considering $\mathcal{O}=1$
such that $\left\langle \mathcal{O}\right\rangle_{\tilde{p}(S; 
    \mathcal{N},q)}=1$ and dividing by $e^{f_q}$ results in
\begin{equation}\label{eq:fs:normalization:constants}
  e^{-f_q} = \sum_{i=1}^{K} \sum_{a=1}^{N_{q_i}}
  \frac{e^{q S_a^{q_i}}}{\sum_{k=1}^{K} N_{q_k}
    e^{q_k S_a^{q_i} + f_{q_k}}}\,.
\end{equation}
While the corresponding constants $f_{q_k}$ in the denominator are determined 
self consistently by setting $f_q = f_{q_j}$ with $j = 1, \dots, K$ on the left 
side of Eq.~\eqref{eq:fs:normalization:constants}.
The values of $f_{q_j}$ can be obtained by iteration, e.g., starting
with $f_{q_j}=0$ for all values of $j$ or by more
sophisticated approaches \cite{bereau2009}. Note that the values of
$f_{q_j}$ can be determined only up to an additive but irrelevant constant.

Reweighting by means of Eq.~\eqref{eq:fs:expectation:value} is
performed over the data points directly, i.e., there is
the additional benefit that it is a histogram-less method. Thus, no 
selection of a specific binning for the measured values of $S$ is necessary.
The calculation of the corresponding statistical error-propagation is 
given in appendix \ref{ap:error:propagation:fs:reweighting},
which, according to the knowledge of the authors,
has not been published in the literature so far.
For only unbiased data samples, a discussion of statistical errors can be 
found in \cite{Rohwer_2015} and for importance-sampling estimators in
\cite{Glasserman_1997}.

With the approach explained above, $\mu(q,\mathcal{N})$ can be obtained 
for basically any desired value of $q$.
Thus, we turn now to the determination
of $\varPsi(q; \mathcal{N})$, which can be written
as $ \frac{1}{\mathcal{N}} \ln
\left\langle e^{q   S} \right\rangle_{p(S; \mathcal{N})}$.
By comparing with Eq.~\eqref{eq:fs:expectation:value} we observe
that the calculation of the average
can be achieved by first setting $q=0$, which also
means that $f_q=f_0$ in the equation, and then using
$\mathcal{O}=e^{qS}$, where now $q$ has become  a free parameter.
The latter yields $\mathcal{O}_a^{q_i} =e^{q S_a^{q_i}}$, when applied to the 
data points.
This results in

\begin{equation}\label{eq:expectation:scgf:finite:scale}
  \left\langle e^{q S}\right\rangle_{p(S; \mathcal{N})} =
  \sum_{i=1}^{K} \sum_{a=1}^{N_{q_i}}
  \frac{e^{q S_a^{q_i} +f_0}}{\sum_{k=1}^{K} N_{q_k}
    e^{q_k S_a^{q_i} + f_{q_k}}}\,.
\end{equation}
While this looks very similar to Eq.~\eqref{eq:fs:expectation:value}
for $\mathcal{O}\equiv 1$, it differs in the fact that $f_0$ appears
in the argument of the exponential, instead of $f_q$. Since the values of
$f_{q_k}$ and $f_0$ are already known, by taking
the logarithm and dividing by $\mathcal{N}$, $\varPsi(s ; \mathcal{N}$
is readily obtained.
It should be mentioned that $\varPsi(q; \mathcal{N})$ could also
be determined from Eq.~\eqref{eq:moments:scgf:derivative} using 
thermodynamic integration \cite{Touchette_2012}, which is not done here
to keep the error-propagation calculations simple for the SCGF estimator.

Since there is data for various scales $\mathcal{N}$, next we want to
 extrapolate $\mu(q; \mathcal{N})$ and $\varPsi(q; \mathcal{N})$ toward 
infinite scales, i.e., taking the limit 
$\mathcal{N}\rightarrow \infty$. For this purpose,
for any choice of  $q$, a function 
$g(\mathcal{N}; q)$ is fitted.
Here, again  power laws
\begin{align}\label{eq:power:law}
  g(\mathcal{N}; q) =& C(q) + A(q) \mathcal{N}^{-B(q)}\\
  &\text{with}\; B(q) \ge 0 \nonumber
\end{align}
are used, where $A$, $B$, and $C$ are $q$ dependent fit-parameters.
The extrapolated value is given by $C(q)$ and is
different when extrapolating $\mu$ or $\Psi$.
The same symbol is only used for simplicity here.
While the true functional form usually depends on the
finite-size behavior at bias parameter $q$ of the considered system and
is unknown 
in general, the power law in Eq. \eqref{eq:power:law} is a robust choice.
For example, there are cases where this form of scaling is known to apply
in a large-deviation context:
In a cloning approach, the corresponding SCGF estimator is shown to follow 
a power-law scaling \cite{Hidalgo_2017, Nemoto_2017}. Note that for
problems which are admissible to 
the cloning approach, the SCGF can be estimated directly from
``growth factors'' describing the evolution of the population,
while our approach is based on Eq.~(\ref{eq:expectation:scgf:finite:scale}), which is very general.
Finite-size corrections to eigenvalue statistics of random matrices were
investigated in \cite{Edelman_2016}, where leading order correction terms 
are found to follow a power-law, too.

The final step is to perform the LF transform in Eq.
\eqref{eq:legendre:fenchel:transformation}.  Instead of fixing
$s$ and then searching for the value of $q$ that maximizes the
supremum argument, it is easier to do a parametric LF
transformation \cite{Touchette_2009, Coghi_2023}: For maximizing
with respect to $q$ one takes
the derivative of the supremum argument and equals it to zero,
which gives
$\varPsi'(q) = s = \mu(q)$.	 Substituting back into
Eq. \eqref{eq:legendre:fenchel:transformation} yields 
\begin{equation}\label{eq:parametric:legendre:transform}
  I(\mu(q)) = q \mu(q) - \varPsi(q).
\end{equation}
This form of the rate function can be directly evaluated
for any desired value of $q$ at
$\mu(q)$ and $\varPsi(q)$, which are the extrapolated values
for $\mathcal{N} \rightarrow \infty$ by means of fitting
Eq.~\eqref{eq:power:law}.  Doing so for sufficiently many values
of $q$, each time yielding $s=\mu(q)$,
will give arbitrary dense resolution on the rate
function $I(s)$.  This is computationally cheap, since
arbitrary values for the bias parameter $q$ can be utilized.
As long as the data samples cover the region of $s$ that is
relevant for the chosen bias $q$, no new simulations are
necessary.  For error-propagation, the correlation between
$\mu(q)$ and $\varPsi(q)$ in
Eq. \eqref{eq:parametric:legendre:transform} is ignored.

\section{Case Studies}
Two systems are analyzed.
First, a slight variation of binomial distributed data, see for example the 
introduction.
This is mostly intended as a proof of concept for the presented method.
The second system is the distribution of the largest connected component in 
ER random graphs, which has a rather complicated
and non-convex rate function in the highly connected phase,
see Eq.~\eqref{eq:rate:function:random:graphs}).
It serves to test the method's performance in a complex system scenario, for 
example in the presence of phase transitions, as will be shown.
ER graphs have been subject to large-deviation studies in various context 
before \cite{Engel_2004, Hartmann_2011, Hartmann_2017, Bhamidi_2015, 
  Coghi_2023}. 
	
\subsection{Binomial Distributed Data}
Revisiting the introductory example,
the probability distribution of $S$ rainy days in a period of
$\mathcal{N}$ days is given by
\begin{equation}\label{eq:binomial:distribution}
  p(S; \mathcal{N}) = \left(
  \begin{array}{c} \mathcal{N}\\ S \end{array}
  \right) r(\mathcal{N})^S (1-r(\mathcal{N}))^{\mathcal{N}-S}.
\end{equation}
Equivalently, also the fraction of rainy days $s=S / \mathcal{N}$ can be
considered, which is done in the following.
To introduce a more challenging scaling behavior, which leads to
a size-dependence of the rate function,
the rain probability
$r(\mathcal{N})$ is set to explicitly have a system scale $\mathcal{N}$ 
dependence of the form
\begin{equation}\label{eq:scale:dependent:probability}
  r(\mathcal{N}):= r_\infty + c \mathcal{N}^{-\gamma},
\end{equation}
where $0 \le r_\infty \le 1$, $\gamma > 0$ and $c$ is a constant.
The parameters have to be chosen such that  $0 \le r(\mathcal{N}) \le 
1$ for all values of $\mathcal{N}$.
The specific choices $r_\infty = 0.25$, $c=7.4$ and
$\gamma=1$ are used here, while the considered system 
scales are $\mathcal{N}_i = \lfloor 10 \cdot(1.25)^i\rfloor$, with $i 
\in \left[ 2, 3, \dots, 17\right]$.

\begin{figure}
  \includegraphics[width=0.8\linewidth]{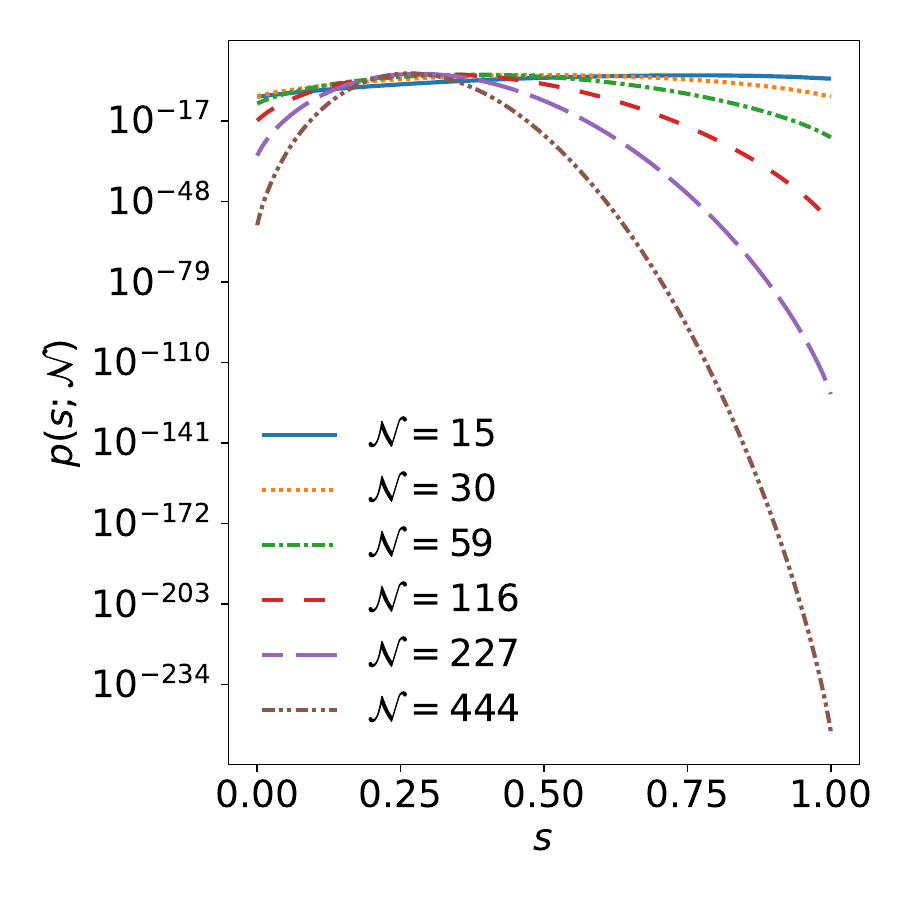}
  \caption{\label{fig:binomial:distributions}
    Normalized binomial distributions with scale dependent parameter 
    $r(\mathcal{N})$ as given in Eq. \eqref{eq:scale:dependent:probability}.
    In contrast to a binomial distribution with constant value of 
    $r(\mathcal{N})$, the mean and variance change with the system
    scale $\mathcal{N}$.
    Only a selection of system scales $\mathcal{N}$ is shown for 
		visual purposes.}
\end{figure}

A plot of the probability distributions in Eq. 
\eqref{eq:binomial:distribution} is shown in figure 
\ref{fig:binomial:distributions} for various scales $\mathcal{N}$.
The curves undergo significant change with increasing scale parameter,
especially the peak of the distribution shifts from around $s\approx 
0.76$ towards $s\approx r_\infty = 0.25$ \\
The reason for making the parameter $r(\mathcal{N})$ explicitly dependent
on the system scale  $\mathcal{N}$ is that this causes the expectation value
\begin{equation}\label{eq:extrapolated:expectation:value}
  \left\langle s\right\rangle =
  \frac{\left\langle S\right\rangle }{\mathcal{N}} =
  r_\infty + c \mathcal{N}^{-\gamma}
\end{equation}
and the SCGF
\begin{equation}\label{eq:cgf:binomial}
  \varPsi(q; \mathcal{N}) = \ln (1-r(\mathcal{N})(1-e^q)),
\end{equation}
here given before taking the limit $\mathcal{N}\rightarrow \infty$,
to be a function of the systems scale $\mathcal{N}$ as well.
Finite-size behavior similar to this, which emerges for most
complex systems naturally without including it explicitly,
can make the estimation of
rate functions difficult. The reason is that simply rescaling the data, i.e.,
dividing it with the scale $\mathcal{N}$, does not necessarily
mitigate finite-size effects.
For this particular case, it means that rescaling the data will
result in a different estimate of the mean
$\left\langle s\right\rangle$ for every scale $\mathcal{N}$, in 
contrast to when $r(\mathcal{N})$ is constant.
	
The scaling behavior in this simple binomial example is known and generally 
follows a power law  due to the specific choice for the parameter 
$r(\mathcal{N})$ in Eq. \eqref{eq:scale:dependent:probability}. 
For the expectation value $\left\langle s\right\rangle$, this can be seen 
directly in Eq. \eqref{eq:extrapolated:expectation:value}.
The SCGF Eq.~(\ref{eq:cgf:binomial}) by Talyor expansion
around $q = 0$,
approximately follows this behavior too, i.e.,
\begin{equation} \label{eq:psi_extrpolation}
  \varPsi(q; \mathcal{N}) \approx q r(\mathcal{N})
  = q r_\infty + q c \mathcal{N}^{-\gamma}\,.
\end{equation}
Therefore, the extrapolation by means of Eq. \eqref{eq:power:law}
is well
justified.

We tested the numerical simulation of biased data, subsequent determination of
$\mu(q, \mathcal{N})$ and $\varPsi(q; \mathcal{N})$, extrapolation
to infinite $\mathcal{N}\to \infty$,
and finally applying the LF to obtain the limiting rate function $I(S)$.
For this purpose, 36 bias values $q$ from the
regime $q \in \left[ -5, \dots , 5\right]$ were selected in a way that
the value ranges close to the peak of the biased distributions cover 
the entire set of $s$ values.

\begin{figure}
  \includegraphics[width=0.8\linewidth]{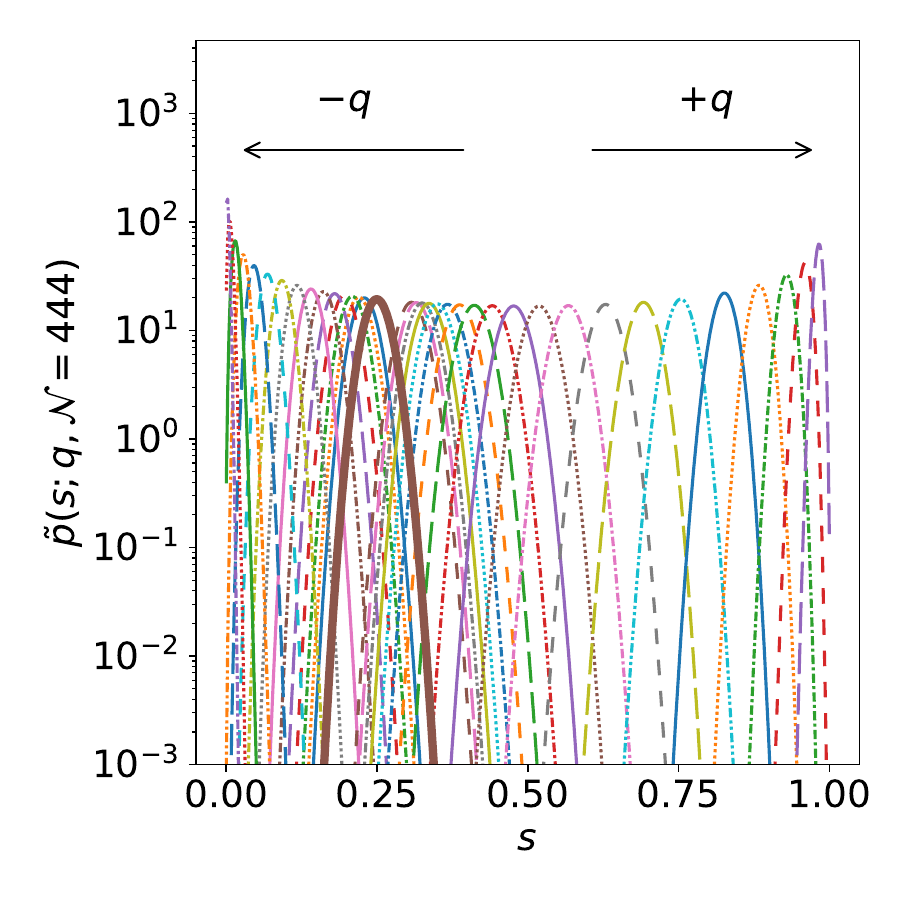}
  \caption{\label{fig:largest:scale:biased:distributions}
    Biased binomial distribution
    $\tilde{p}(s; \mathcal{N}, q) \sim p(s;\mathcal{N}) \exp(q s)$
    at system scale $\mathcal{N} = 444$ for the various bias values $q$.
    The thicker curve shows the unbiased case, i.e., $q=0$.
    Depending on the specific bias value $q$, the distributions are 
    shifted towards higher or smaller values of the fraction of rainy 
    days $s$.
    Each distributions peak region stretches over a different subset of 
    $s$, such that in total the entire range of $s$ is covered.}
\end{figure}
\begin{figure}
	\includegraphics[width=0.8\linewidth]{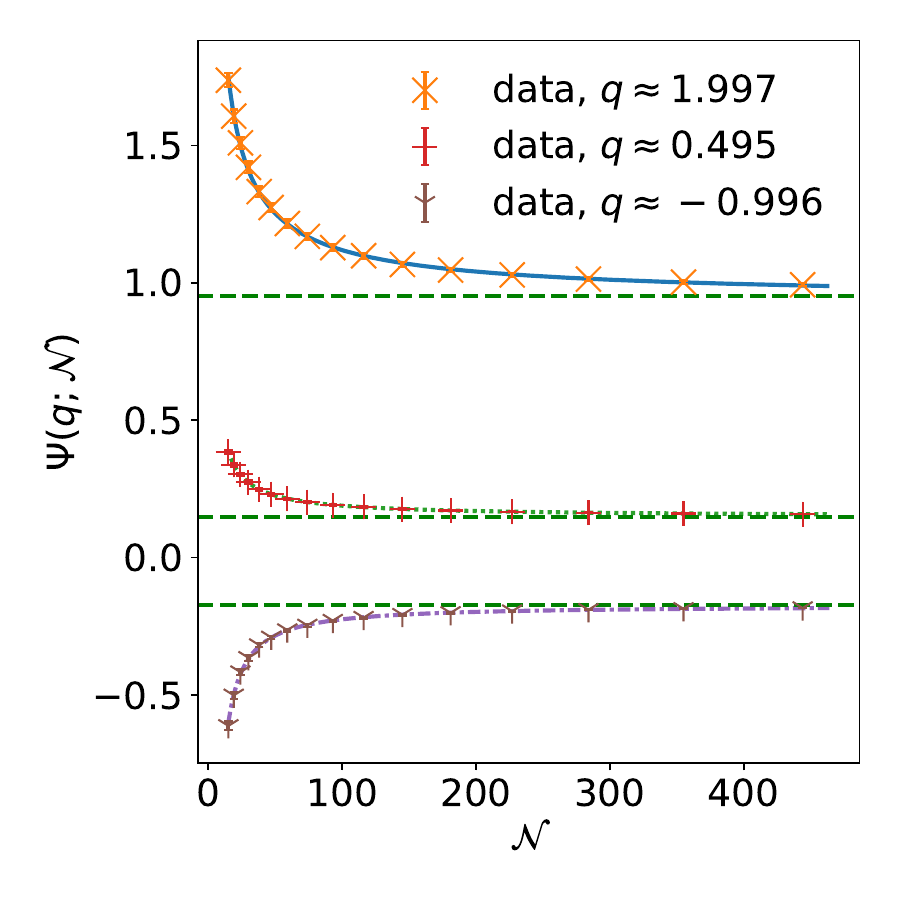}
	\includegraphics[width=0.8\linewidth]{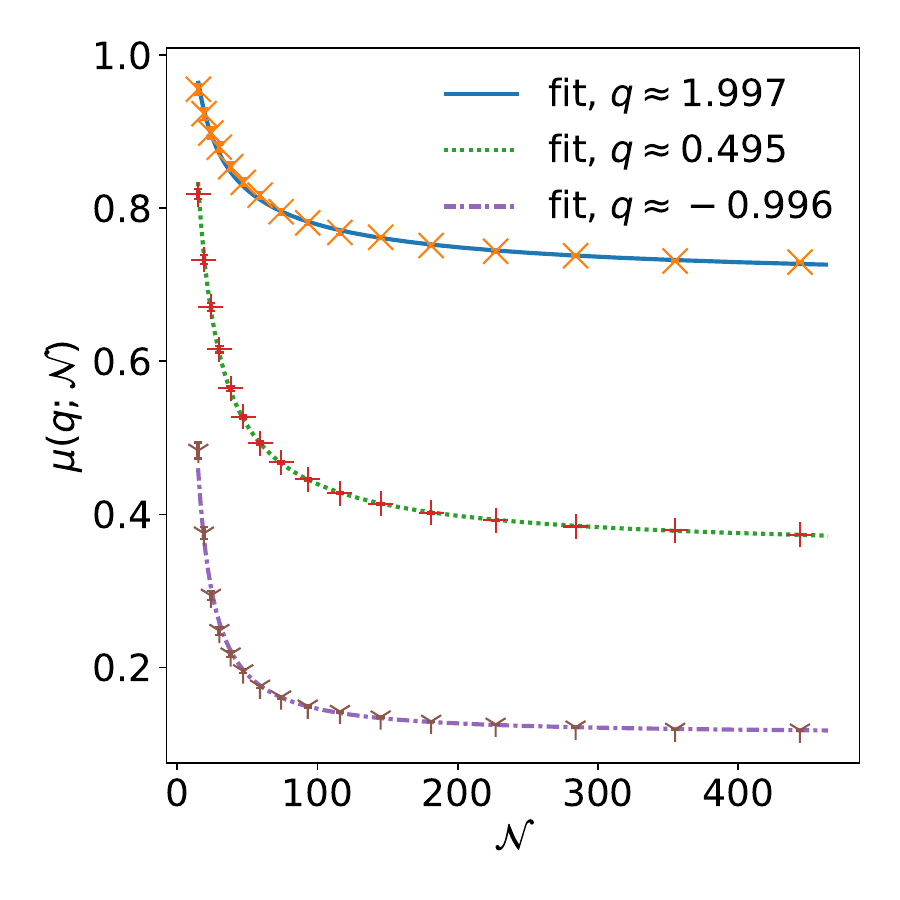}
	\caption{\label{fig:extraploation:SCGF:and:mean:binomial:case}
		Fit of the power-law in Eq. \eqref{eq:power:law} for the binomial 
		distributed	data to extrapolate towards infinite system sizes
		$\mathcal{N}$.
		Curves for a selection of bias values $q$ are shown together with the
		corresponding fits.
		Top: SCGF $\varPsi(q; \mathcal{N})$.
		Also shown are the correct limiting values according to Eq. 
		\eqref{eq:cgf:binomial} as dashed horizontal green lines.
		Bottom: Biased expectation value $\mu(q; \mathcal{N})$.}
\end{figure}

An example of the resulting biased probability distributions 
$\tilde{p}(s; \mathcal{N}, q) \sim p(s;\mathcal{N}) \exp(q \mathcal{N} s)$
is depicted in figure \ref{fig:largest:scale:biased:distributions}
for the largest system scale $\mathcal{N}=444$, where
the unbiased case, i.e. $q=0$, is highlighted with a thicker line.
The system scales used for the binomial case are in fact so small
that direct sampling is possible, i.e., calculating all probabilities
$\tilde{p}(s; \mathcal{N}, q)$ explicitly and then draw for each sample
the number $S$ of rainy days accordingly.
For every bias value $q$, $N= 3 \cdot \mathcal{N}$ data points are 
generated, thus a very small number as compared to typical number
of samples needed for statistical analyzes.
Exemplary extrapolation fits for the SCGF and the expectation 
value at various biases parameters $q$ are shown in figure
\ref{fig:extraploation:SCGF:and:mean:binomial:case}.
The power law behavior is found to match the data decently.

\begin{figure}
  \includegraphics[width=0.8\linewidth]{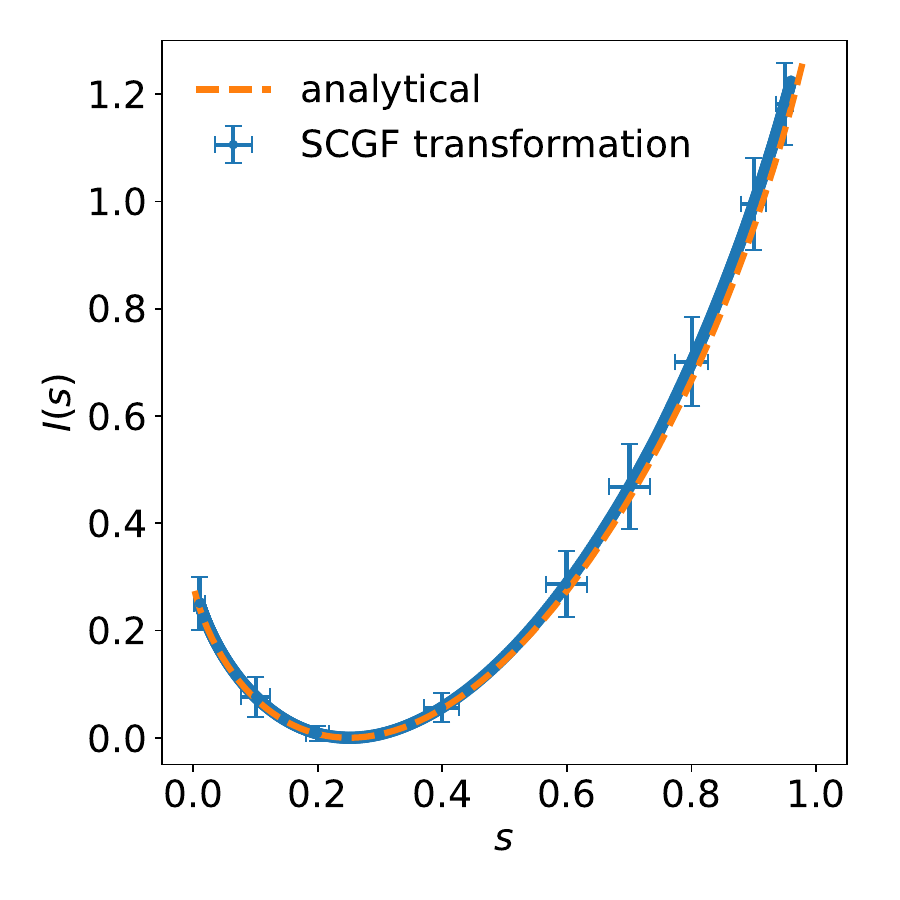}
  \caption{\label{fig:binomial:rate:function}
    Comparison of the exact analytical rate function in Eq. 
    \eqref{eq:rate:function:binomial} and the rate function
    obtained from the parametric Legendre-Fenchel transformation 
    (see Eq. \eqref{eq:parametric:legendre:transform}) of 
    the numerically estimated SCGF for the number of rainy days.
    For visual purposes only a representative selection of errorbars
    are shown.}
\end{figure}

In figure \ref{fig:binomial:rate:function} the
parametric LF transformation (see Eq. 
\eqref{eq:parametric:legendre:transform}) of the extrapolated SCGF is
depicted. 
This is compared to the analytical rate function 
\begin{equation}\label{eq:rate:function:binomial}
  I(s) = s \ln \frac{s}{r_\infty} + (1-s) \ln 
  \frac{1-s}{1-r_\infty}\,,
\end{equation}
which is identical to the form when $r(\mathcal{N})$ is scale independent
\cite{Peliti_2021}.
Note that due the parametric LF transformation also the 
horizontal axis for $s$ exhibits error bars.
A good agreement is found within errors over the entire support of the rate
function $I(s)$.

\subsection{Largest Connected Component of Erd\"os-R\'enyi Random Graphs}
As second case study, the largest connected component size of ER random 
graphs \cite{Engel_2004} is considered.
Each graph consists of $\mathcal{N}$ nodes. For each
pair $i,j$ of nodes an edge is randomly
and independently inserted with probability $c/(\mathcal{N}-1)$, where the
connectivity $c$ denotes the  mean number of edges that are
incident to a node. This graph ensemble exhibits a percolation transition
at $c=1$. For $c<1$ the graph is composed of many small tree-like
components, where for $c>1$ there exists in the limit $\mathcal{N}\to\infty$
one large component, which is of the order of the system size.

The analytical rate function for the relative size of the largest connected 
component $s=S/\mathcal{N}$ is known \cite{OConnell_1998}.
It is given piecewise on intervals $[s_k, s_{k-1}] \subset [0,1]$
which are arranged from right to left:
\begin{align}\label{eq:rate:function:random:graphs}
  s_0 = &1 \nonumber\\
  s_k = &\sup_s \left\lbrace \frac{s}{1-ks}=1-e^{-cs}\right\rbrace 
  \nonumber \\
  m(y) = &\log\left(1-e^{-y}\right) \nonumber \\
  I_(s) = &-ksm(cs) + ks \log(s) + (1-ks)\log(1-ks) 
  \nonumber \\
  &+ cks - k(k+1) cs^2/2 \nonumber
  \nonumber \\
  &\text{for} \quad s_k < s \le s_{k-1},
\end{align}
Note that for $c\le 1$, $s_1=0$, thus there is only one interval $[s_1,s_0]$,
where for $c>1$ in principle infinitely intervals are needed to describe
$I(s)$, but almost all of them accumulate near $s=0$.

Here, the two representative
connectivities $c \in \left\lbrace 0.5, 2.0 \right\rbrace$
for graphs of order $\mathcal{N} \in \left\lbrace50,\\75,\\100,
\\150,\\200,\\300,\\400\right\rbrace$ were investigated.
We considered 34 bias values $q \in [-3.85, \dots, 3.85]$,
where one is always the unbiased case, i.e., $q=0$.
For sampling, the same methods as described in \cite{Hartmann_2011} are 
used, $10^3$ independent data points were generated for each value of $q$.
        
\begin{figure}[ht]
  \includegraphics[width=0.8\linewidth]{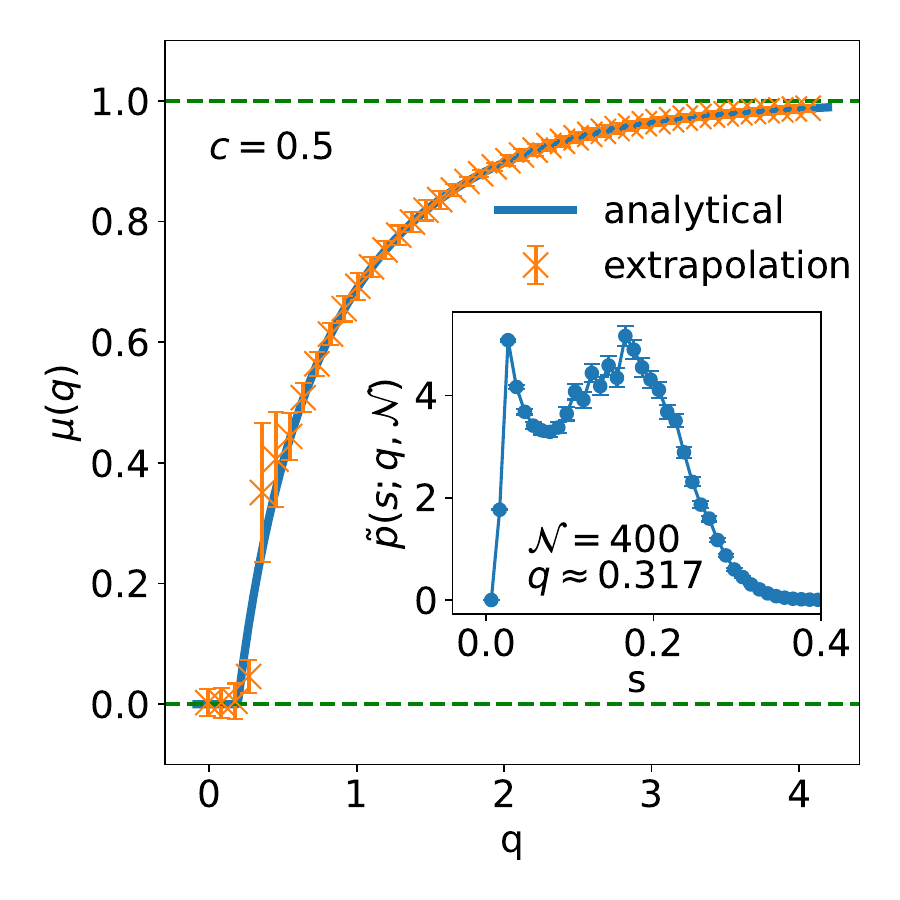}
  \caption{\label{fig:random:graphs:c05:first:moment:vs:q}
    For connectivity $c=0.5$, analytical and extrapolated relative size 
    of the larges connected	component $\left\langle s\right\rangle$ of 
    the Erd\"os R\'enyi random graph ensemble.
    The extrapolation is according to Eq. \eqref{eq:power:law}.
    The dashed horizontal lines indicate the limiting
    values of the expectation value $\left\langle s\right\rangle$
    for $q\rightarrow \pm \infty$.
    The gap at $q\approx 0.3$ for the extrapolated curve is caused by a 
    phase transition in the corresponding biased ensemble, 
    resulting in deviations from the assumed power law scaling 
    behavior.
    This transition is also indicated by the double peak structure in the 
    distribution of the largest connected component relative size for graph 
    order $\mathcal{N}=400$ and bias $q\approx0.317$ that is shown in the inset.
	The lines in the inset are a guide to the eye only.
	}
\end{figure}

For connectivity $c=0.5$, figure 
\ref{fig:random:graphs:c05:first:moment:vs:q} shows the mean relative
size of the largest connected component $\left\langle  s \right\rangle $ as 
a function of the bias value $q$.
The analytical curve is obtained numerically from the rate function in
Eq. \eqref{eq:rate:function:random:graphs}, which means first performing
a LF transformation of $I(s)$ and then calculating the 
derivative, since $\varPsi'(q) = \left\langle s \right\rangle$.
Except for a region around $q\approx 0.3$, the analytical result is
reproduced within error by the extrapolation via Eq. \eqref{eq:power:law}.

\begin{figure}[ht]
	\includegraphics[width=0.8\linewidth]{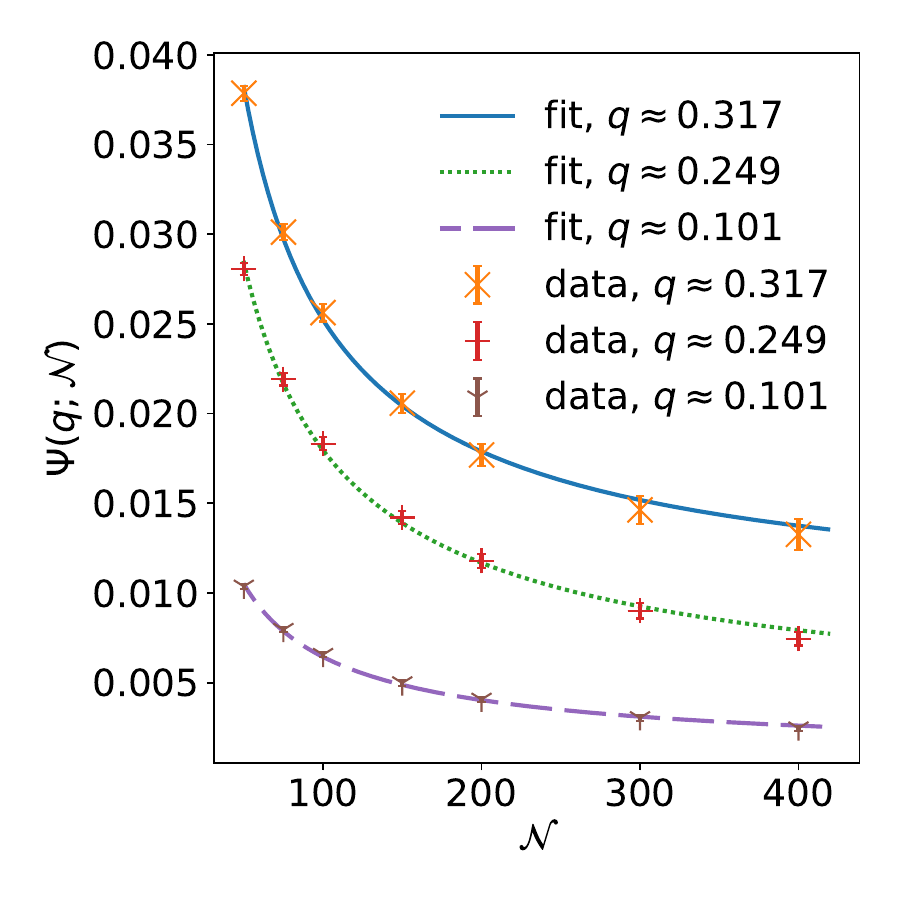}
	\includegraphics[width=0.8\linewidth]{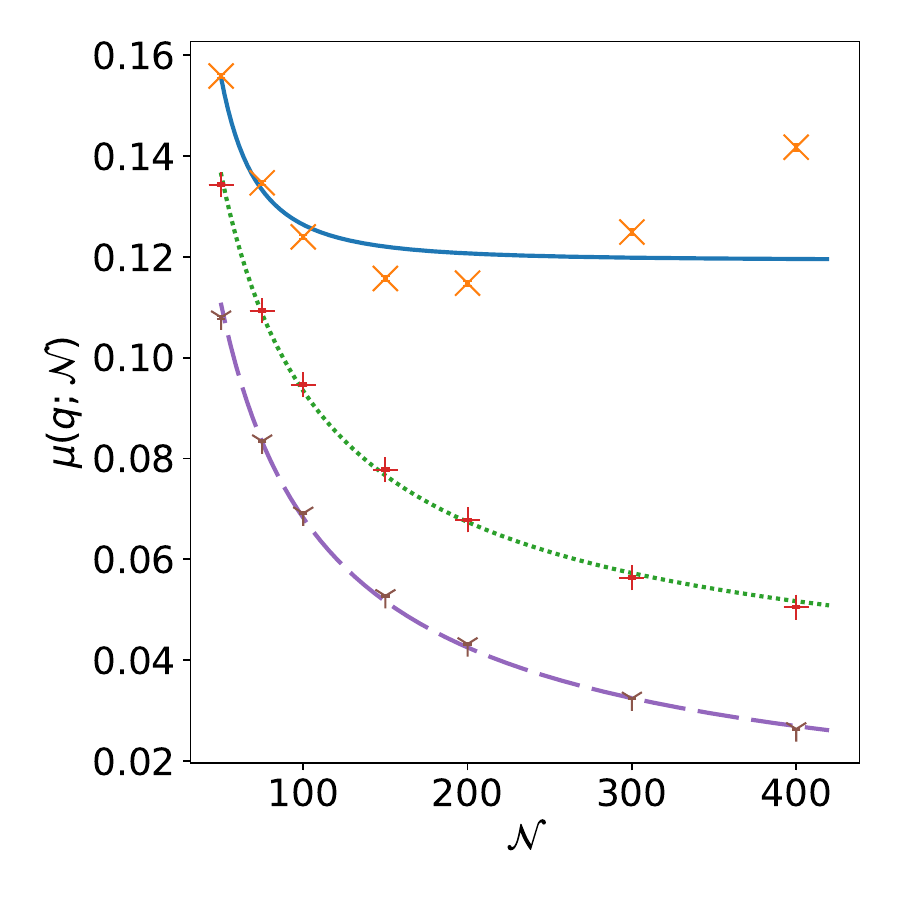}
	\caption{\label{fig:extraploation:SCGF:and:mean:random:graphs}
		Infinite system size $\mathcal{N}$ extrapolation fit by means of the power-law in Eq. \eqref{eq:power:law} for the largest connected component in the Erd\"os-R\'enyi random graph ensemble with connectivity $c=0.5$.
		Curves for a selection of bias values $q$ are shown together with the
		corresponding fits.
		Top: SCGF $\varPsi(q; \mathcal{N})$.
		Bottom: Biased expectation value $\mu(q; \mathcal{N})$.
		The scaling behavior for $q\approx 3.17$ shows substantial
		deviations from that of a power law.}
\end{figure}

At this value of $q$, the biased ensemble experiences a phase transition as 
can be seen from the double peak structure in the corresponding probability 
distribution of the largest component size, which is shown in the inset of 
figure \ref{fig:random:graphs:c05:first:moment:vs:q}.
This causes a different scaling behavior, making the
extrapolation by means of a power law inaccurate around this bias value
regime (see figure \ref{fig:extraploation:SCGF:and:mean:random:graphs}).
A similar break-down of scaling in presence of a phase transition was 
observed in \cite{Hidalgo_2018}, too.

\begin{figure}
  \includegraphics[width=0.8\linewidth]{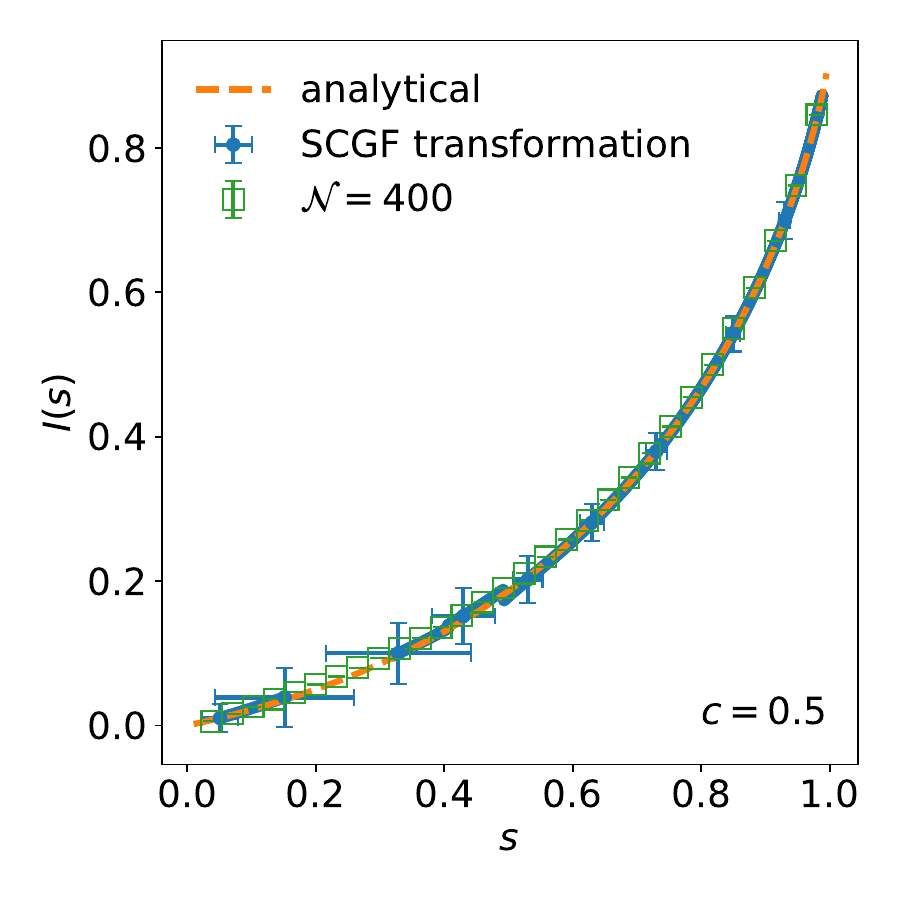}
  \caption{\label{fig:random:graphs:rate:function:c05}
    Comparison of the analytical rate function in Eq. 
    \eqref{eq:rate:function:random:graphs}, the Legendre-Fenchel 
    transformed extrapolated SCGF and the empirical rate function at 
    system size $\mathcal{N}=400$ for the Erd\"os R\'enyi random graph
    ensemble.
    The connectivity is $c=0.5$.
    The gap in the transformed SCGF curve at $s\approx 0.25$ is due to 
    the phase transition of the biased graph ensemble at this point and
    the resulting breakdown of the assumed scaling power law in Eq.
    \eqref{eq:power:law} used for extrapolation.
    For visual purposes, only a representative selection of errorbars
    are shown on the curve obtained by Legendre-Fenchel transformation
    of the estimated SCGF.}
\end{figure}

In figure \ref{fig:random:graphs:rate:function:c05}, the exact rate 
function, the LF transformed size-extrapolated SCGF and the empirical rate
function Eq. \eqref{eq:empirical:rate:function} for the largest
graph at $\mathcal{N}=400$ are
shown for $c=0.5$
All curves have good agreement within errors. In the  
regime around $s\approx0.25$ in the transformed SCGF curve a larger
error bars appear that are due to the aforementioned phase transition.
Also very close to $s\approx0.25$, $I(s)$ could not be estimated by
the present approach.
	
\begin{figure}
  \includegraphics[width=0.8\linewidth]{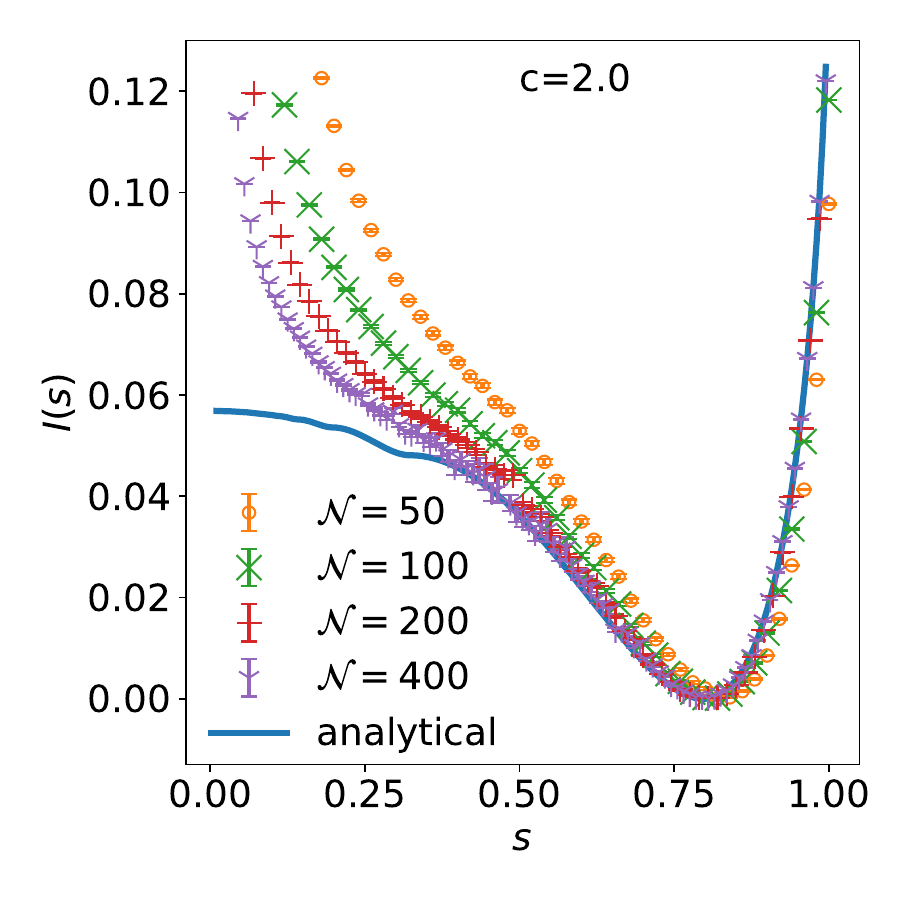}
  \caption{\label{fig:random:graphs:c2:empirical:rate:func}
    For the Erd\"os R\'enyi random graph
    ensemble at connectivity $c=2.0$, the analytical rate function in 
    Eq. \eqref{eq:rate:function:random:graphs} and empirical rate 
    functions at various system sales $\mathcal{N}$.
    While the analytical curve is already matched for values $s\geq0.8$
    by the curves at finite system sizes $\mathcal{N}$, the convergence 
    at smaller values of $s$ is slower.}
\end{figure}

For connectivity $c=2.0$, the rate function has concave regions and its 
minimum is no longer located at $s=0$, as can be seen in 
figure \ref{fig:random:graphs:c2:empirical:rate:func}.
Both, the analytical solution and empirical rate functions at various 
scales $\mathcal{N}$ are depicted.
Convergence of the empirical rate function towards the true curve is
observed for increasing graph order $\mathcal{N}$, while the convergence
is slower for smaller values of $s$.

\begin{figure}
  \includegraphics[width=0.8\linewidth]{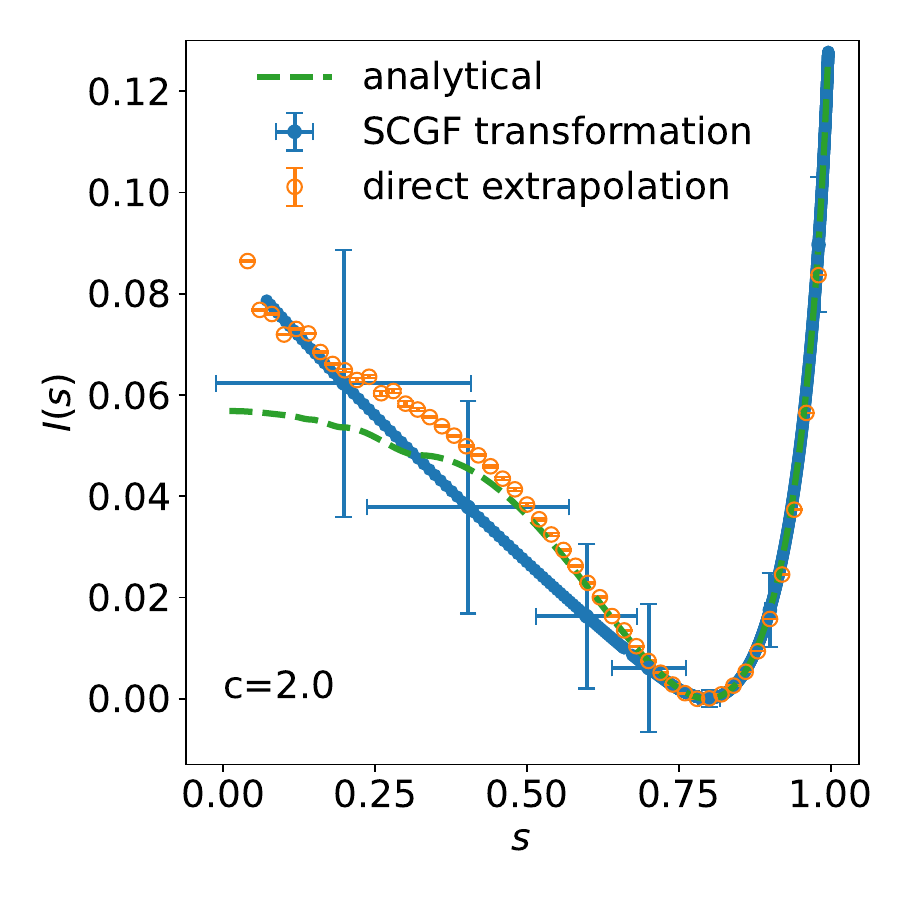}
  \caption{\label{fig:random:graphs:c2:estimation:comparision}
    Plot of the analytical rate function, the directly extrapolated rate
    function and the  Legendre-Fenchel transformed SCGF.
    The transformed SCGF only yields the convex envelope of the
    true rate function in the regime $s \approx 0.3 \dots 0.75$.
    There the analytical rate function is concave.
    For visual purposes, only a representative selection of errorbars
    are shown on the curve obtained by Legendre-Fenchel transformation
    of the estimated SCGF.}
\end{figure}

Here, in addition to obtain $I(s)$ by our approach from extrapolating
and transforming the SCGF, we also directly
extrapolated the numerically obtained \cite{Hartmann_2011}
finite-size rate functions. This was achieved by fitting
Eq.~\eqref{eq:power:law:s:space} as a function of $\mathcal{N}$, for every 
of the 34 values of $s$ where we obtained data. This
yields the curve denoted as ``direct extrapolation'' shown in figure 
\ref{fig:random:graphs:c2:estimation:comparision}.
Concave parts of the rate function can still be 
reproduced by this method, like in the region around $s\approx 0.6$, while
the LF transformed SCGF can only yield the 
convex envelope in such regimes, which is indeed the case but with a
a rather large statistical error estimate.
Note that the LF transformed SCGF has arbitrary spacial
resolution, even if only systems at finite scales were simulated, in 
contrast to the directly extrapolated rate function.
The convex region, i.e., $s\geq 0.75$, is reproduced by both methods, while
for values $s < 0.3$ no accurate estimate could be achieved by any of the
considered methods.
	
\section{Discussion}
We presented an approach to estimate large-deviation rate functions.
The method is based on biased sampling of the quantity of interest, 
multi-data histogram-less reweighting, followed by  with a
finite-size extrapolation 
fit of the SCGF and the tilted expectation value, with a final application
of the Legendre-Fenchel transform.
Also an analysis of the resulting statistical
errors was performed, as shown in the appendix.
The intended targets of the approach are models which exhibit strong
finite-size behavior in 
the context of numerical simulations, specifically when applying
importance-sampling approaches.

Two example systems with known analytical solutions were considered:
First, a modified binomial distributed variable, for
which the rate function could be reliably reproduced.
Second, the distribution of the size of the
largest connected component  for ER graphs.
For the low connectivity case, a phase transition in the biased system 
distribution changed the finite-size behavior such that the extrapolation
fit had bad quality, which led to a gap in the estimated rate function.
In the high connectivity phase, the rate function could be reproduced in
the convex regions, otherwise only yielding the convex envelope.
Due to this non-convexity at very small values of the relative size of the
largest connected 
component, which exhibited very slow convergence with increasing system 
scale, the method could not outperform other methods.
Still, for this case
the obtained statistical errors are rather large, such that the
known analytical result is actually included within error bars.
On the other hand, the direct extrapolation of the rate function exhibits
very small error bars, such that for $s\le 0.3$ the direct extrapolation
is statistically not comptabile with analytic exact result.

It could be interesting to test different models with entirely convex rate 
functions, but strong finite-size effects, in the future.
This could further reveal strengths and shortcomings of the method.
	
\begin{acknowledgments}
  The authors wish to thank Mylène Maïda for pointing out the useful
  reference \cite{Edelman_2016} on finite-size corrections of the eigenvalue 
  statistics of random matrices.
\end{acknowledgments}
	
	\appendix
	
	\section{Tilted Expectation Value}\label{ap:biased:moments}
	Using the exponentially biased distribution
	\begin{equation}\label{eq:biased:distribution}
		\tilde{p}(S; \mathcal{N}, q) :=
		\frac{e^{q S}}{\mathcal{Z}(q,\mathcal{N})} p(S; \mathcal{N}),
	\end{equation}
	where $\mathcal{Z}(q,\mathcal{N})$ is a normalization constant, the 
	following relation
	\begin{align}\label{eq:derivation1:biased:moments}
		\left\langle S e^{q S} \right\rangle_{p(S; \mathcal{N})} 
		&:=\int_{-\infty}^{\infty} S e^{q S}
		p(S; \mathcal{N})dS \nonumber \\
		&=\mathcal{Z}(q,\mathcal{N}) \int_{-\infty}^{\infty}
		S \tilde{p}(S; \mathcal{N}, q)dS \nonumber \\
		&=\mathcal{Z}(q, \mathcal{N}) \left\langle S
		\right\rangle_{\tilde{p}(S; \mathcal{N}, q)}
	\end{align}
	is obtained.
	Inserting this result into the tilted expectation value yields
	\begin{align}\label{eq:derivation2:biased:moments}
		\mu(q; \mathcal{N}) &= \frac{\left\langle S 
		e^{q S} \right\rangle_{p(S; \mathcal{N})}}{\mathcal{N}
		\left\langle e^{q S} \right\rangle_{p(S; \mathcal{N})}} 
		\nonumber\\
	 	&=\frac{\mathcal{Z}(q, \mathcal{N}) \left\langle S
	 	\right\rangle_{\tilde{p}(S;\mathcal{N},q)}}
 		{\mathcal{N}\mathcal{Z}(q,
	 	\mathcal{N}) \left\langle 1 
	 	\right\rangle_{\tilde{p}(S; \mathcal{N}, q)}} \nonumber\\
 	 	&= \frac{\left\langle S \right\rangle_{\tilde{p}(S; 
 	 			\mathcal{N}, q)}}{\mathcal{N}},
	\end{align}
	which is exactly Eq. \eqref{eq:biased:moments}.
	
\section{Error-propagation in Multiple Histogram-Reweighting}
\label{ap:error:propagation:fs:reweighting}
The variance of the observable $S$ simulated at bias value $q_i$
is estimated by
\begin{equation} \label{eq:var:S}
  \left(\sigma^{q_i}_S\right)^2 = \frac{1}{N-1} \sum_{a=1}^{N_{q_i}}
  \left(S_a^{q_i} - \frac{1}{N_{q_i}} \sum_{b=1}^{N_{q_i}} 
  S_b^{q_i}\right)^2.
\end{equation}
To obtain the corresponding variance for the expectation value $\mu$
and the SCGF $\varPsi$, Gaussian error propagation is performed.
This means, for a function $g(x_1,\ldots,x_n)$,
where the random numbers $x_1,\ldots, x_n$ exhibit variances
$\sigma_1^2,\ldots, \sigma_n^2$ and are uncorrelated,
the total variance is estimated
by $\sum_i \left(\frac{\partial g}{\partial x_i}\right)^2\sigma_i^2$.

In both cases $\mu$ and $\varPsi$,
the derivative of the normalization constants $f_q$,
defined in Eq. \eqref{eq:fs:normalization:constants}, with
respect to a particular data point $S_a^{q_l}$ ($a=1,\ldots,N_{q_l}$) is required, 
which is given by
\begin{multline}\label{eq:normalization:derivative}
  \frac{\partial f_q}{\partial S_a^{q_l}} =
  -\frac{q e^{q S_a^{q_l} +f_q}}{\sum_{i=1}^{K} N_i e^{q_i 
      S_a^{q_l} + f_i}} \\
  + e^{q S_a^{q_l} +f_q} \frac{\sum_{j=1}^{K} 
    N_{q_j} q_j e^{q_j S_a^{q_l} + f_{q_j}}}
  {\left(\sum_{i=1}^{K} N_{q_i}   e^{q_i S_a^{q_l} + f_{q_i}} \right)^2 }\\
  + \sum_{j=1}^{K} \sum_{k=1}^{K} \sum_{b=1}^{N_i}
  \frac{N_{q_j} e^{q_j S_b^{q_k} + f_{q_j}} e^{q S_b^{q_k} + 
      f_q}}{\left( \sum_{i=1}^{K} N_{q_i}
    e^{q_i S_b^{q_k} + f_{q_i}}\right)^2} \frac{\partial 
    f_{q_j}}{\partial S_a^{q_l}}.
\end{multline}
The necessary derivatives $\frac{\partial f_{q_j}}{\partial S_a^{q_l}}$ 
are determined by setting $f_q = f_{q_i}$.
This turns Eq. \eqref{eq:normalization:derivative} into a set of linear
equation that needs to be solved for every data point.

\paragraph{Expectation Value}
The expectation value through reweighting, see Eq. 
\eqref{eq:fs:expectation:value}, at a fixed value of $q$ is given by
\begin{equation}
  \mu(q;\mathcal{N}) =
  \sum_{i=1}^{K} \sum_{a=1}^{N_{q_i}}
  \frac{S_a^{q_i} e^{q S_a^{q_i} +f_q}}{\sum_{k=1}^{K} N_{q_k}
    e^{q_k S_a^{q_i} + f_{q_k}}},
\end{equation}
with the normalization constants $f_{q_k}$ from
Eq. \eqref{eq:fs:normalization:constants}.
The corresponding variance is then calculated via
\begin{equation}
  \left(\sigma_{\mu(q;\mathcal{N})}\right)^2 = \sum_{i=1}^{K} 
  \sum_{a=1}^{N_{q_i}}
  \left( \frac{\partial\mu(q;\mathcal{N})}{\partial S_a^{q_i}} 
  \right)^2 \left(\sigma^{q_i}_S\right)^2,
\end{equation}
where we naturally
assume here and below that all sampled values $S_a^{q_i}$ exhibit the
same variance $\left(\sigma^{q_i}_S\right)^2$ given
by Eq.~(\ref{eq:var:S}). The needed
partial derivatives read as 
\begin{multline}
  \frac{\partial \mu(q; \mathcal{N})}{\partial S_a^{q_l}} = 
  \frac{(1 + q S_a^{q_l})
    e^{q S_a^{q_l} +f_q}}{\sum_{i=1}^{K} N_{q_i} e^{q_i 
      S_a^{q_l} + f_{q_i}}} \\
  - S_a^{q_l} e^{q S_a^{q_l} +f_q} \frac{\sum_{j=1}^{K} 
    N_{q_j} q_j e^{q_j S_a^{q_l} + f_{q_j}}}{\left(\sum_{i=1}^{K} N_{q_i}
    e^{q_i S_a^{q_l} + f_{q_i}} \right)^2 } \\
  + \mu(q; \mathcal{N}) \frac{\partial f_q}{\partial S_a^{q_l}} \\
  - \sum_{j=1}^{K} \sum_{k=1}^{K} \sum_{b=1}^{N_{q_k}}
  \frac{N_{q_j} S_b^{q_k} e^{q_j S_b^{q_k} + f_{q_j}} e^{q 
      S_b^{q_k} + f_q}}{\left( \sum_{i=1}^{K} N_{q_i}
    e^{q_i S_b^{q_k} + f_{q_i}}\right)^2}
  \frac{\partial f_{q_j}}{\partial S_a^{q_l}}\,.
\end{multline}

\paragraph{SCGF}
The error-propagation for the SCGF
\begin{equation}\label{eq:finite:scale:scgf}
  \varPsi(q; \mathcal{N}) :=
  \frac{1}{\mathcal{N}} \ln \left\langle e^{q S} 
  \right\rangle_{p(S; \mathcal{N})},
\end{equation}
at a finite scale $\mathcal{N}$ and fixed bias parameter $q$,
goes in a similar fashion.
Resolving the angular brackets in Eq. \eqref{eq:finite:scale:scgf}
through Eq. \eqref{eq:fs:expectation:value}, as shown above, yields
Eq.~(\ref{eq:expectation:scgf:finite:scale}).
The variance for the SCGF is then
\begin{multline}
  \left(\sigma_{\varPsi(q; \mathcal{N})}\right)^2 = \sum_{i=1}^{K} 
  \sum_{a=1}^{N_{q_i}}
  \left( \frac{\partial\varPsi(q; 
    \mathcal{N})}{\partial S_a^{q_i}} \right)^2
  \left(\sigma^{q_i}_S\right)^2\\
  = \frac{1}{\left( \mathcal{N}\left\langle  e^{q S} 
    \right\rangle_{p(S; \mathcal{N})}\right)^2}
  \sum_{i=1}^{K} \sum_{a=1}^{N_{q_i}}
  \left( \frac{\partial \left\langle  e^{q S} 
    \right\rangle_{p(S; \mathcal{N})}}{\partial S_a^{q_i}} 
  \right)^2
  \left(\sigma^{q_i}_S\right)^2,
\end{multline}
with derivatives
\begin{multline}
  \frac{\partial \left\langle e^{q S}\right\rangle_{p(S; 
      \mathcal{N})}}{\partial S_a^{q_l}} =
  + \left\langle e^{q S}\right\rangle_{p(S; \mathcal{N})} 
  \frac{\partial f_q}{\partial S_a^{q_l}}\\	
  +\frac{q e^{q S_a^{q_l} +f_0}}{\sum_{i=1}^{K} N_{q_i} e^{q_i 
      S_a^{q_l} + f_{q_i}}} \\	
  - e^{q S_a^{q_l} +f_0} \frac{\sum_{j=1}^{K} 
    N_{q_j} q_j e^{q_j S_a^{q_l} + f_{q_j}}}{\left(\sum_{i=1}^{K} N_{q_i}
    e^{q_i S_a^{q_l} + f_{q_i}} \right)^2 }\\
  - \sum_{j=1}^{K} \sum_{k=1}^{K} \sum_{b=1}^{N_{q_k}}
  \frac{N_{q_j} e^{q_j S_b^{q_k} + f_{q_j}} e^{q S_b^{q_k} + f_0}}
       {\left( \sum_{i=1}^{K} N_{q_i}e^{q_i S_b^{q_k} + f_{q_i}}\right)^2}
  \frac{\partial   f_{q_j}}{\partial S_a^{q_l}}.
\end{multline}
	
	\bibliography{references} % Produces the bibliography via BibTeX.

%merlin.mbs apsrev4-1.bst 2010-07-25 4.21a (PWD, AO, DPC) hacked
%Control: key (0)
%Control: author (8) initials jnrlst
%Control: editor formatted (1) identically to author
%Control: production of article title (-1) disabled
%Control: page (0) single
%Control: year (1) truncated
%Control: production of eprint (0) enabled
\begin{thebibliography}{47}%
\makeatletter
\providecommand \@ifxundefined [1]{%
 \@ifx{#1\undefined}
}%
\providecommand \@ifnum [1]{%
 \ifnum #1\expandafter \@firstoftwo
 \else \expandafter \@secondoftwo
 \fi
}%
\providecommand \@ifx [1]{%
 \ifx #1\expandafter \@firstoftwo
 \else \expandafter \@secondoftwo
 \fi
}%
\providecommand \natexlab [1]{#1}%
\providecommand \enquote  [1]{``#1''}%
\providecommand \bibnamefont  [1]{#1}%
\providecommand \bibfnamefont [1]{#1}%
\providecommand \citenamefont [1]{#1}%
\providecommand \href@noop [0]{\@secondoftwo}%
\providecommand \href [0]{\begingroup \@sanitize@url \@href}%
\providecommand \@href[1]{\@@startlink{#1}\@@href}%
\providecommand \@@href[1]{\endgroup#1\@@endlink}%
\providecommand \@sanitize@url [0]{\catcode `\\12\catcode `\$12\catcode
  `\&12\catcode `\#12\catcode `\^12\catcode `\_12\catcode `\%12\relax}%
\providecommand \@@startlink[1]{}%
\providecommand \@@endlink[0]{}%
\providecommand \url  [0]{\begingroup\@sanitize@url \@url }%
\providecommand \@url [1]{\endgroup\@href {#1}{\urlprefix }}%
\providecommand \urlprefix  [0]{URL }%
\providecommand \Eprint [0]{\href }%
\providecommand \doibase [0]{http://dx.doi.org/}%
\providecommand \selectlanguage [0]{\@gobble}%
\providecommand \bibinfo  [0]{\@secondoftwo}%
\providecommand \bibfield  [0]{\@secondoftwo}%
\providecommand \translation [1]{[#1]}%
\providecommand \BibitemOpen [0]{}%
\providecommand \bibitemStop [0]{}%
\providecommand \bibitemNoStop [0]{.\EOS\space}%
\providecommand \EOS [0]{\spacefactor3000\relax}%
\providecommand \BibitemShut  [1]{\csname bibitem#1\endcsname}%
\let\auto@bib@innerbib\@empty
%</preamble>
\bibitem [{\citenamefont {den Hollander}(2000)}]{denHollander2000}%
  \BibitemOpen
  \bibfield  {author} {\bibinfo {author} {\bibfnamefont {F.}~\bibnamefont {den
  Hollander}},\ }\href@noop {} {\emph {\bibinfo {title} {Large Deviations}}}\
  (\bibinfo  {publisher} {American Mathematical Society},\ \bibinfo {address}
  {Providence},\ \bibinfo {year} {2000})\BibitemShut {NoStop}%
\bibitem [{\citenamefont {Touchette}(2009)}]{Touchette_2009}%
  \BibitemOpen
  \bibfield  {author} {\bibinfo {author} {\bibfnamefont {H.}~\bibnamefont
  {Touchette}},\ }\href {\doibase
  https://doi.org/10.1016/j.physrep.2009.05.002} {\bibfield  {journal}
  {\bibinfo  {journal} {Physics Reports}\ }\textbf {\bibinfo {volume} {478}},\
  \bibinfo {pages} {1} (\bibinfo {year} {2009})}\BibitemShut {NoStop}%
\bibitem [{\citenamefont {Touchette}(2012)}]{Touchette_2012}%
  \BibitemOpen
  \bibfield  {author} {\bibinfo {author} {\bibfnamefont {H.}~\bibnamefont
  {Touchette}},\ }\href {https://arxiv.org/abs/1106.4146} {\enquote {\bibinfo
  {title} {A basic introduction to large deviations: Theory, applications,
  simulations},}\ } (\bibinfo {year} {2012}),\ \Eprint
  {http://arxiv.org/abs/1106.4146} {arXiv:1106.4146 [cond-mat.stat-mech]}
  \BibitemShut {NoStop}%
\bibitem [{\citenamefont {Engel}\ \emph {et~al.}(2004)\citenamefont {Engel},
  \citenamefont {Monasson},\ and\ \citenamefont {Hartmann}}]{Engel_2004}%
  \BibitemOpen
  \bibfield  {author} {\bibinfo {author} {\bibfnamefont {A.}~\bibnamefont
  {Engel}}, \bibinfo {author} {\bibfnamefont {R.}~\bibnamefont {Monasson}}, \
  and\ \bibinfo {author} {\bibfnamefont {A.~K.}\ \bibnamefont {Hartmann}},\
  }\href {\doibase 10.1007/s10955-004-2268-6} {\bibfield  {journal} {\bibinfo
  {journal} {Journal of Statistical Physics}\ }\textbf {\bibinfo {volume}
  {117}},\ \bibinfo {pages} {387} (\bibinfo {year} {2004})}\BibitemShut
  {NoStop}%
\bibitem [{\citenamefont {Hartmann}(2011)}]{Hartmann_2011}%
  \BibitemOpen
  \bibfield  {author} {\bibinfo {author} {\bibfnamefont {A.~K.}\ \bibnamefont
  {Hartmann}},\ }\href {\doibase 10.1140/epjb/e2011-10836-4} {\bibfield
  {journal} {\bibinfo  {journal} {Eur. Phys. J. B}\ }\textbf {\bibinfo {volume}
  {84}},\ \bibinfo {pages} {627} (\bibinfo {year} {2011})}\BibitemShut
  {NoStop}%
\bibitem [{\citenamefont {Hartmann}(2017)}]{Hartmann_2017}%
  \BibitemOpen
  \bibfield  {author} {\bibinfo {author} {\bibfnamefont {A.~K.}\ \bibnamefont
  {Hartmann}},\ }\href {\doibase 10.1140/epjst/e2016-60368-3} {\bibfield
  {journal} {\bibinfo  {journal} {The European Physical Journal Special
  Topics}\ }\textbf {\bibinfo {volume} {226}},\ \bibinfo {pages} {567}
  (\bibinfo {year} {2017})}\BibitemShut {NoStop}%
\bibitem [{\citenamefont {Giardin\`a}\ \emph {et~al.}(2006)\citenamefont
  {Giardin\`a}, \citenamefont {Kurchan},\ and\ \citenamefont
  {Peliti}}]{Giardina_2006}%
  \BibitemOpen
  \bibfield  {author} {\bibinfo {author} {\bibfnamefont {C.}~\bibnamefont
  {Giardin\`a}}, \bibinfo {author} {\bibfnamefont {J.}~\bibnamefont {Kurchan}},
  \ and\ \bibinfo {author} {\bibfnamefont {L.}~\bibnamefont {Peliti}},\ }\href
  {\doibase 10.1103/PhysRevLett.96.120603} {\bibfield  {journal} {\bibinfo
  {journal} {Phys. Rev. Lett.}\ }\textbf {\bibinfo {volume} {96}},\ \bibinfo
  {pages} {120603} (\bibinfo {year} {2006})}\BibitemShut {NoStop}%
\bibitem [{\citenamefont {Lecomte}\ and\ \citenamefont
  {Tailleur}(2007)}]{Lecomte_2007}%
  \BibitemOpen
  \bibfield  {author} {\bibinfo {author} {\bibfnamefont {V.}~\bibnamefont
  {Lecomte}}\ and\ \bibinfo {author} {\bibfnamefont {J.}~\bibnamefont
  {Tailleur}},\ }\href {\doibase 10.1088/1742-5468/2007/03/P03004} {\bibfield
  {journal} {\bibinfo  {journal} {Journal of Statistical Mechanics: Theory and
  Experiment}\ }\textbf {\bibinfo {volume} {2007}},\ \bibinfo {pages} {P03004}
  (\bibinfo {year} {2007})}\BibitemShut {NoStop}%
\bibitem [{\citenamefont {Guevara~Hidalgo}\ \emph {et~al.}(2017)\citenamefont
  {Guevara~Hidalgo}, \citenamefont {Nemoto},\ and\ \citenamefont
  {Lecomte}}]{Hidalgo_2017}%
  \BibitemOpen
  \bibfield  {author} {\bibinfo {author} {\bibfnamefont {E.}~\bibnamefont
  {Guevara~Hidalgo}}, \bibinfo {author} {\bibfnamefont {T.}~\bibnamefont
  {Nemoto}}, \ and\ \bibinfo {author} {\bibfnamefont {V.}~\bibnamefont
  {Lecomte}},\ }\href {\doibase 10.1103/PhysRevE.95.062134} {\bibfield
  {journal} {\bibinfo  {journal} {Phys. Rev. E}\ }\textbf {\bibinfo {volume}
  {95}},\ \bibinfo {pages} {062134} (\bibinfo {year} {2017})}\BibitemShut
  {NoStop}%
\bibitem [{\citenamefont {K\"orner}\ \emph {et~al.}(2006)\citenamefont
  {K\"orner}, \citenamefont {Katzgraber},\ and\ \citenamefont
  {Hartmann}}]{pe_sk2006}%
  \BibitemOpen
  \bibfield  {author} {\bibinfo {author} {\bibfnamefont {M.}~\bibnamefont
  {K\"orner}}, \bibinfo {author} {\bibfnamefont {H.~G.}\ \bibnamefont
  {Katzgraber}}, \ and\ \bibinfo {author} {\bibfnamefont {A.~K.}\ \bibnamefont
  {Hartmann}},\ }\href {\doibase 10.1088/1742-5468/2006/04/P04005} {\bibfield
  {journal} {\bibinfo  {journal} {J. Stat. Mech.}\ }\textbf {\bibinfo {volume}
  {2006}},\ \bibinfo {pages} {P04005} (\bibinfo {year} {2006})}\BibitemShut
  {NoStop}%
\bibitem [{\citenamefont {Cérou}\ and\ \citenamefont
  {Guyader}(2007)}]{Cerou_2007}%
  \BibitemOpen
  \bibfield  {author} {\bibinfo {author} {\bibfnamefont {F.}~\bibnamefont
  {Cérou}}\ and\ \bibinfo {author} {\bibfnamefont {A.}~\bibnamefont
  {Guyader}},\ }\href {\doibase 10.1080/07362990601139628} {\bibfield
  {journal} {\bibinfo  {journal} {Stochastic Analysis and Applications}\
  }\textbf {\bibinfo {volume} {25}},\ \bibinfo {pages} {417} (\bibinfo {year}
  {2007})},\ \Eprint
  {http://arxiv.org/abs/https://doi.org/10.1080/07362990601139628}
  {https://doi.org/10.1080/07362990601139628} \BibitemShut {NoStop}%
\bibitem [{\citenamefont {Agranov}\ \emph {et~al.}(2020)\citenamefont
  {Agranov}, \citenamefont {Zilber}, \citenamefont {Smith}, \citenamefont
  {Admon}, \citenamefont {Roichman},\ and\ \citenamefont
  {Meerson}}]{Agranov_2020}%
  \BibitemOpen
  \bibfield  {author} {\bibinfo {author} {\bibfnamefont {T.}~\bibnamefont
  {Agranov}}, \bibinfo {author} {\bibfnamefont {P.}~\bibnamefont {Zilber}},
  \bibinfo {author} {\bibfnamefont {N.~R.}\ \bibnamefont {Smith}}, \bibinfo
  {author} {\bibfnamefont {T.}~\bibnamefont {Admon}}, \bibinfo {author}
  {\bibfnamefont {Y.}~\bibnamefont {Roichman}}, \ and\ \bibinfo {author}
  {\bibfnamefont {B.}~\bibnamefont {Meerson}},\ }\href {\doibase
  10.1103/PhysRevResearch.2.013174} {\bibfield  {journal} {\bibinfo  {journal}
  {Phys. Rev. Res.}\ }\textbf {\bibinfo {volume} {2}},\ \bibinfo {pages}
  {013174} (\bibinfo {year} {2020})}\BibitemShut {NoStop}%
\bibitem [{\citenamefont {Hartmann}\ \emph {et~al.}(2013)\citenamefont
  {Hartmann}, \citenamefont {Majumdar},\ and\ \citenamefont
  {Rosso}}]{fBm_MC2013}%
  \BibitemOpen
  \bibfield  {author} {\bibinfo {author} {\bibfnamefont {A.~K.}\ \bibnamefont
  {Hartmann}}, \bibinfo {author} {\bibfnamefont {S.~N.}\ \bibnamefont
  {Majumdar}}, \ and\ \bibinfo {author} {\bibfnamefont {A.}~\bibnamefont
  {Rosso}},\ }\href {\doibase 10.1103/PhysRevE.88.022119} {\bibfield  {journal}
  {\bibinfo  {journal} {Phys. Rev. E}\ }\textbf {\bibinfo {volume} {88}},\
  \bibinfo {pages} {022119} (\bibinfo {year} {2013})}\BibitemShut {NoStop}%
\bibitem [{\citenamefont {Hartmann}\ and\ \citenamefont
  {Meerson}(2024)}]{fBm_PA2024}%
  \BibitemOpen
  \bibfield  {author} {\bibinfo {author} {\bibfnamefont {A.~K.}\ \bibnamefont
  {Hartmann}}\ and\ \bibinfo {author} {\bibfnamefont {B.}~\bibnamefont
  {Meerson}},\ }\href {\doibase 10.1103/PhysRevE.109.014146} {\bibfield
  {journal} {\bibinfo  {journal} {Phys. Rev. E}\ }\textbf {\bibinfo {volume}
  {109}},\ \bibinfo {pages} {014146} (\bibinfo {year} {2024})}\BibitemShut
  {NoStop}%
\bibitem [{\citenamefont {Smith}\ and\ \citenamefont
  {Majumdar}(2022)}]{Smith_2022}%
  \BibitemOpen
  \bibfield  {author} {\bibinfo {author} {\bibfnamefont {N.~R.}\ \bibnamefont
  {Smith}}\ and\ \bibinfo {author} {\bibfnamefont {S.~N.}\ \bibnamefont
  {Majumdar}},\ }\href {\doibase 10.1088/1742-5468/ac6f04} {\bibfield
  {journal} {\bibinfo  {journal} {Journal of Statistical Mechanics: Theory and
  Experiment}\ }\textbf {\bibinfo {volume} {2022}},\ \bibinfo {pages} {053212}
  (\bibinfo {year} {2022})}\BibitemShut {NoStop}%
\bibitem [{\citenamefont {Staffeldt}\ and\ \citenamefont
  {Hartmann}(2019)}]{nagel_schreckenberg2019}%
  \BibitemOpen
  \bibfield  {author} {\bibinfo {author} {\bibfnamefont {W.}~\bibnamefont
  {Staffeldt}}\ and\ \bibinfo {author} {\bibfnamefont {A.~K.}\ \bibnamefont
  {Hartmann}},\ }\href {\doibase 10.1103/PhysRevE.100.062301} {\bibfield
  {journal} {\bibinfo  {journal} {Phys. Rev. E}\ }\textbf {\bibinfo {volume}
  {100}},\ \bibinfo {pages} {062301} (\bibinfo {year} {2019})}\BibitemShut
  {NoStop}%
\bibitem [{\citenamefont {Hartmann}\ \emph {et~al.}(2018)\citenamefont
  {Hartmann}, \citenamefont {Doussal}, \citenamefont {Majumdar}, \citenamefont
  {Rosso},\ and\ \citenamefont {Schehr}}]{kpz2018}%
  \BibitemOpen
  \bibfield  {author} {\bibinfo {author} {\bibfnamefont {A.~K.}\ \bibnamefont
  {Hartmann}}, \bibinfo {author} {\bibfnamefont {P.~L.}\ \bibnamefont
  {Doussal}}, \bibinfo {author} {\bibfnamefont {S.~N.}\ \bibnamefont
  {Majumdar}}, \bibinfo {author} {\bibfnamefont {A.}~\bibnamefont {Rosso}}, \
  and\ \bibinfo {author} {\bibfnamefont {G.}~\bibnamefont {Schehr}},\ }\href
  {\doibase 10.1209/0295-5075/121/67004} {\bibfield  {journal} {\bibinfo
  {journal} {Europhys. Lett.}\ }\textbf {\bibinfo {volume} {121}},\ \bibinfo
  {pages} {67004} (\bibinfo {year} {2018})}\BibitemShut {NoStop}%
\bibitem [{\citenamefont {Hartmann}\ \emph
  {et~al.}(2019{\natexlab{a}})\citenamefont {Hartmann}, \citenamefont
  {Meerson},\ and\ \citenamefont {Sasorov}}]{kpz_HMS2019}%
  \BibitemOpen
  \bibfield  {author} {\bibinfo {author} {\bibfnamefont {A.~K.}\ \bibnamefont
  {Hartmann}}, \bibinfo {author} {\bibfnamefont {B.}~\bibnamefont {Meerson}}, \
  and\ \bibinfo {author} {\bibfnamefont {P.}~\bibnamefont {Sasorov}},\ }\href
  {\doibase 10.1103/PhysRevResearch.1.032043} {\bibfield  {journal} {\bibinfo
  {journal} {Phys. Rev. Research}\ }\textbf {\bibinfo {volume} {1}},\ \bibinfo
  {pages} {032043} (\bibinfo {year} {2019}{\natexlab{a}})}\BibitemShut
  {NoStop}%
\bibitem [{\citenamefont {Hartmann}\ \emph {et~al.}(2020)\citenamefont
  {Hartmann}, \citenamefont {Krajenbrink},\ and\ \citenamefont
  {Le~Doussal}}]{kpz_long2020}%
  \BibitemOpen
  \bibfield  {author} {\bibinfo {author} {\bibfnamefont {A.~K.}\ \bibnamefont
  {Hartmann}}, \bibinfo {author} {\bibfnamefont {A.}~\bibnamefont
  {Krajenbrink}}, \ and\ \bibinfo {author} {\bibfnamefont {P.}~\bibnamefont
  {Le~Doussal}},\ }\href {\doibase 10.1103/PhysRevE.101.012134} {\bibfield
  {journal} {\bibinfo  {journal} {Phys. Rev. E}\ }\textbf {\bibinfo {volume}
  {101}},\ \bibinfo {pages} {012134} (\bibinfo {year} {2020})}\BibitemShut
  {NoStop}%
\bibitem [{\citenamefont {Smith}(2022)}]{Smith_chaotic_maps_2022}%
  \BibitemOpen
  \bibfield  {author} {\bibinfo {author} {\bibfnamefont {N.~R.}\ \bibnamefont
  {Smith}},\ }\href {\doibase 10.1103/PhysRevE.106.L042202} {\bibfield
  {journal} {\bibinfo  {journal} {Phys. Rev. E}\ }\textbf {\bibinfo {volume}
  {106}},\ \bibinfo {pages} {L042202} (\bibinfo {year} {2022})}\BibitemShut
  {NoStop}%
\bibitem [{\citenamefont {Hartmann}(2014)}]{Hartmann_2014}%
  \BibitemOpen
  \bibfield  {author} {\bibinfo {author} {\bibfnamefont {A.~K.}\ \bibnamefont
  {Hartmann}},\ }\href {\doibase 10.1103/PhysRevE.89.052103} {\bibfield
  {journal} {\bibinfo  {journal} {Phys. Rev. E}\ }\textbf {\bibinfo {volume}
  {89}},\ \bibinfo {pages} {052103} (\bibinfo {year} {2014})}\BibitemShut
  {NoStop}%
\bibitem [{\citenamefont {Werner}\ and\ \citenamefont
  {Hartmann}(2021)}]{Werner_2021}%
  \BibitemOpen
  \bibfield  {author} {\bibinfo {author} {\bibfnamefont {P.}~\bibnamefont
  {Werner}}\ and\ \bibinfo {author} {\bibfnamefont {A.~K.}\ \bibnamefont
  {Hartmann}},\ }\href {\doibase 10.1103/PhysRevE.104.034407} {\bibfield
  {journal} {\bibinfo  {journal} {Phys. Rev. E}\ }\textbf {\bibinfo {volume}
  {104}},\ \bibinfo {pages} {034407} (\bibinfo {year} {2021})}\BibitemShut
  {NoStop}%
\bibitem [{\citenamefont {Werner}\ \emph {et~al.}(2024)\citenamefont {Werner},
  \citenamefont {Hartmann},\ and\ \citenamefont {Majumdar}}]{Werner_2024}%
  \BibitemOpen
  \bibfield  {author} {\bibinfo {author} {\bibfnamefont {P.}~\bibnamefont
  {Werner}}, \bibinfo {author} {\bibfnamefont {A.~K.}\ \bibnamefont
  {Hartmann}}, \ and\ \bibinfo {author} {\bibfnamefont {S.~N.}\ \bibnamefont
  {Majumdar}},\ }\href {\doibase 10.1103/PhysRevE.110.024115} {\bibfield
  {journal} {\bibinfo  {journal} {Phys. Rev. E}\ }\textbf {\bibinfo {volume}
  {110}},\ \bibinfo {pages} {024115} (\bibinfo {year} {2024})}\BibitemShut
  {NoStop}%
\bibitem [{\citenamefont {Bucklew}(2004)}]{bucklew2004}%
  \BibitemOpen
  \bibfield  {author} {\bibinfo {author} {\bibfnamefont {J.~A.}\ \bibnamefont
  {Bucklew}},\ }\href@noop {} {\emph {\bibinfo {title} {Introduction to rare
  event simulation}}}\ (\bibinfo  {publisher} {Springer-Verlag},\ \bibinfo
  {address} {New York},\ \bibinfo {year} {2004})\BibitemShut {NoStop}%
\bibitem [{\citenamefont {Bouchet}\ \emph {et~al.}(2019)\citenamefont
  {Bouchet}, \citenamefont {Rolland},\ and\ \citenamefont
  {Wouters}}]{bouchet2019}%
  \BibitemOpen
  \bibfield  {author} {\bibinfo {author} {\bibfnamefont {F.}~\bibnamefont
  {Bouchet}}, \bibinfo {author} {\bibfnamefont {J.}~\bibnamefont {Rolland}}, \
  and\ \bibinfo {author} {\bibfnamefont {J.}~\bibnamefont {Wouters}},\ }\href
  {\doibase 10.1063/1.5120509} {\bibfield  {journal} {\bibinfo  {journal}
  {Chaos}\ }\textbf {\bibinfo {volume} {29}},\ \bibinfo {pages} {080402}
  (\bibinfo {year} {2019})}\BibitemShut {NoStop}%
\bibitem [{\citenamefont {Tailleur}\ and\ \citenamefont
  {Lecomte}(2009)}]{Tailleur_2009}%
  \BibitemOpen
  \bibfield  {author} {\bibinfo {author} {\bibfnamefont {J.}~\bibnamefont
  {Tailleur}}\ and\ \bibinfo {author} {\bibfnamefont {V.}~\bibnamefont
  {Lecomte}},\ }\href {\doibase 10.1063/1.3082284} {\bibfield  {journal}
  {\bibinfo  {journal} {AIP Conference Proceedings}\ }\textbf {\bibinfo
  {volume} {1091}},\ \bibinfo {pages} {212} (\bibinfo {year} {2009})},\ \Eprint
  {http://arxiv.org/abs/https://pubs.aip.org/aip/acp/article-pdf/1091/1/212/11954017/212\_1\_online.pdf}
  {https://pubs.aip.org/aip/acp/article-pdf/1091/1/212/11954017/212\_1\_online.pdf}
  \BibitemShut {NoStop}%
\bibitem [{\citenamefont {Bréhier}(2015)}]{Brehier_2015}%
  \BibitemOpen
  \bibfield  {author} {\bibinfo {author} {\bibfnamefont {C.-E.}\ \bibnamefont
  {Bréhier}},\ }\href@noop {} {\bibfield  {journal} {\bibinfo  {journal}
  {Latin American journal of probability and mathematical statistics}\ }\textbf
  {\bibinfo {volume} {12}} (\bibinfo {year} {2015})}\BibitemShut {NoStop}%
\bibitem [{\citenamefont {Cérou}\ \emph {et~al.}(2019)\citenamefont {Cérou},
  \citenamefont {Guyader},\ and\ \citenamefont {Rousset}}]{Cerou_2019}%
  \BibitemOpen
  \bibfield  {author} {\bibinfo {author} {\bibfnamefont {F.}~\bibnamefont
  {Cérou}}, \bibinfo {author} {\bibfnamefont {A.}~\bibnamefont {Guyader}}, \
  and\ \bibinfo {author} {\bibfnamefont {M.}~\bibnamefont {Rousset}},\ }\href
  {\doibase 10.1063/1.5082247} {\bibfield  {journal} {\bibinfo  {journal}
  {Chaos: An Interdisciplinary Journal of Nonlinear Science}\ }\textbf
  {\bibinfo {volume} {29}},\ \bibinfo {pages} {043108} (\bibinfo {year}
  {2019})},\ \Eprint
  {http://arxiv.org/abs/https://pubs.aip.org/aip/cha/article-pdf/doi/10.1063/1.5082247/14620716/043108\_1\_online.pdf}
  {https://pubs.aip.org/aip/cha/article-pdf/doi/10.1063/1.5082247/14620716/043108\_1\_online.pdf}
  \BibitemShut {NoStop}%
\bibitem [{\citenamefont {Coghi}\ and\ \citenamefont
  {Touchette}(2023)}]{Coghi_2023}%
  \BibitemOpen
  \bibfield  {author} {\bibinfo {author} {\bibfnamefont {F.}~\bibnamefont
  {Coghi}}\ and\ \bibinfo {author} {\bibfnamefont {H.}~\bibnamefont
  {Touchette}},\ }\href {\doibase 10.1103/PhysRevE.107.034137} {\bibfield
  {journal} {\bibinfo  {journal} {Phys. Rev. E}\ }\textbf {\bibinfo {volume}
  {107}},\ \bibinfo {pages} {034137} (\bibinfo {year} {2023})}\BibitemShut
  {NoStop}%
\bibitem [{\citenamefont {Hartmann}\ \emph
  {et~al.}(2019{\natexlab{b}})\citenamefont {Hartmann}, \citenamefont {Kebiri},
  \citenamefont {Neureither},\ and\ \citenamefont {Richter}}]{HartmannC_2019}%
  \BibitemOpen
  \bibfield  {author} {\bibinfo {author} {\bibfnamefont {C.}~\bibnamefont
  {Hartmann}}, \bibinfo {author} {\bibfnamefont {O.}~\bibnamefont {Kebiri}},
  \bibinfo {author} {\bibfnamefont {L.}~\bibnamefont {Neureither}}, \ and\
  \bibinfo {author} {\bibfnamefont {L.}~\bibnamefont {Richter}},\ }\href
  {\doibase 10.1063/1.5090271} {\bibfield  {journal} {\bibinfo  {journal}
  {Chaos: An Interdisciplinary Journal of Nonlinear Science}\ }\textbf
  {\bibinfo {volume} {29}},\ \bibinfo {pages} {063107} (\bibinfo {year}
  {2019}{\natexlab{b}})},\ \Eprint
  {http://arxiv.org/abs/https://pubs.aip.org/aip/cha/article-pdf/doi/10.1063/1.5090271/13534861/063107\_1\_online.pdf}
  {https://pubs.aip.org/aip/cha/article-pdf/doi/10.1063/1.5090271/13534861/063107\_1\_online.pdf}
  \BibitemShut {NoStop}%
\bibitem [{\citenamefont {Grafke}\ and\ \citenamefont
  {Vanden-Eijnden}(2019)}]{Grafke_2019}%
  \BibitemOpen
  \bibfield  {author} {\bibinfo {author} {\bibfnamefont {T.}~\bibnamefont
  {Grafke}}\ and\ \bibinfo {author} {\bibfnamefont {E.}~\bibnamefont
  {Vanden-Eijnden}},\ }\href {\doibase 10.1063/1.5084025} {\bibfield  {journal}
  {\bibinfo  {journal} {Chaos}\ }\textbf {\bibinfo {volume} {29}},\ \bibinfo
  {pages} {063118} (\bibinfo {year} {2019})},\ \Eprint
  {http://arxiv.org/abs/https://pubs.aip.org/aip/cha/article-pdf/doi/10.1063/1.5084025/13536944/063118\_1\_online.pdf}
  {https://pubs.aip.org/aip/cha/article-pdf/doi/10.1063/1.5084025/13536944/063118\_1\_online.pdf}
  \BibitemShut {NoStop}%
\bibitem [{\citenamefont {Alqahtani}\ and\ \citenamefont
  {Grafke}(2021)}]{Alqahtani_2021}%
  \BibitemOpen
  \bibfield  {author} {\bibinfo {author} {\bibfnamefont {M.}~\bibnamefont
  {Alqahtani}}\ and\ \bibinfo {author} {\bibfnamefont {T.}~\bibnamefont
  {Grafke}},\ }\href {\doibase 10.1088/1751-8121/abe67b} {\bibfield  {journal}
  {\bibinfo  {journal} {Journal of Physics A: Mathematical and Theoretical}\
  }\textbf {\bibinfo {volume} {54}},\ \bibinfo {pages} {175001} (\bibinfo
  {year} {2021})}\BibitemShut {NoStop}%
\bibitem [{\citenamefont {Glasserman}\ and\ \citenamefont
  {Wang}(1997)}]{Glasserman_1997}%
  \BibitemOpen
  \bibfield  {author} {\bibinfo {author} {\bibfnamefont {P.}~\bibnamefont
  {Glasserman}}\ and\ \bibinfo {author} {\bibfnamefont {Y.}~\bibnamefont
  {Wang}},\ }\href {http://www.jstor.org/stable/2245293} {\bibfield  {journal}
  {\bibinfo  {journal} {The Annals of Applied Probability}\ }\textbf {\bibinfo
  {volume} {7}},\ \bibinfo {pages} {731} (\bibinfo {year} {1997})}\BibitemShut
  {NoStop}%
\bibitem [{\citenamefont {Ferrenberg}\ and\ \citenamefont
  {Swendsen}(1988)}]{Ferrenberg_1988}%
  \BibitemOpen
  \bibfield  {author} {\bibinfo {author} {\bibfnamefont {A.~M.}\ \bibnamefont
  {Ferrenberg}}\ and\ \bibinfo {author} {\bibfnamefont {R.~H.}\ \bibnamefont
  {Swendsen}},\ }\href {\doibase 10.1103/PhysRevLett.61.2635} {\bibfield
  {journal} {\bibinfo  {journal} {Phys. Rev. Lett.}\ }\textbf {\bibinfo
  {volume} {61}},\ \bibinfo {pages} {2635} (\bibinfo {year}
  {1988})}\BibitemShut {NoStop}%
\bibitem [{\citenamefont {Ferrenberg}\ and\ \citenamefont
  {Swendsen}(1989)}]{Ferrenberg_1989}%
  \BibitemOpen
  \bibfield  {author} {\bibinfo {author} {\bibfnamefont {A.~M.}\ \bibnamefont
  {Ferrenberg}}\ and\ \bibinfo {author} {\bibfnamefont {R.~H.}\ \bibnamefont
  {Swendsen}},\ }\href {\doibase 10.1103/PhysRevLett.63.1195} {\bibfield
  {journal} {\bibinfo  {journal} {Phys. Rev. Lett.}\ }\textbf {\bibinfo
  {volume} {63}},\ \bibinfo {pages} {1195} (\bibinfo {year}
  {1989})}\BibitemShut {NoStop}%
\bibitem [{\citenamefont {Kumar}\ \emph {et~al.}(1992)\citenamefont {Kumar},
  \citenamefont {Rosenberg}, \citenamefont {Bouzida}, \citenamefont
  {Swendsen},\ and\ \citenamefont {Kollman}}]{kumar1992}%
  \BibitemOpen
  \bibfield  {author} {\bibinfo {author} {\bibfnamefont {S.}~\bibnamefont
  {Kumar}}, \bibinfo {author} {\bibfnamefont {J.~M.}\ \bibnamefont
  {Rosenberg}}, \bibinfo {author} {\bibfnamefont {D.}~\bibnamefont {Bouzida}},
  \bibinfo {author} {\bibfnamefont {R.~H.}\ \bibnamefont {Swendsen}}, \ and\
  \bibinfo {author} {\bibfnamefont {P.~A.}\ \bibnamefont {Kollman}},\ }\href
  {\doibase https://doi.org/10.1002/jcc.540130812} {\bibfield  {journal}
  {\bibinfo  {journal} {J. Comp. Chem.}\ }\textbf {\bibinfo {volume} {13}},\
  \bibinfo {pages} {1011} (\bibinfo {year} {1992})}\BibitemShut {NoStop}%
\bibitem [{\citenamefont {Ferrenberg}\ \emph {et~al.}(1995)\citenamefont
  {Ferrenberg}, \citenamefont {Landau},\ and\ \citenamefont
  {Swendsen}}]{Ferrenberg_1995}%
  \BibitemOpen
  \bibfield  {author} {\bibinfo {author} {\bibfnamefont {A.~M.}\ \bibnamefont
  {Ferrenberg}}, \bibinfo {author} {\bibfnamefont {D.~P.}\ \bibnamefont
  {Landau}}, \ and\ \bibinfo {author} {\bibfnamefont {R.~H.}\ \bibnamefont
  {Swendsen}},\ }\href {\doibase 10.1103/PhysRevE.51.5092} {\bibfield
  {journal} {\bibinfo  {journal} {Phys. Rev. E}\ }\textbf {\bibinfo {volume}
  {51}},\ \bibinfo {pages} {5092} (\bibinfo {year} {1995})}\BibitemShut
  {NoStop}%
\bibitem [{\citenamefont {Bereau}\ and\ \citenamefont
  {Swendsen}(2009)}]{bereau2009}%
  \BibitemOpen
  \bibfield  {author} {\bibinfo {author} {\bibfnamefont {T.}~\bibnamefont
  {Bereau}}\ and\ \bibinfo {author} {\bibfnamefont {R.~H.}\ \bibnamefont
  {Swendsen}},\ }\href {\doibase https://doi.org/10.1016/j.jcp.2009.05.011}
  {\bibfield  {journal} {\bibinfo  {journal} {J. Comp. Phys.}\ }\textbf
  {\bibinfo {volume} {228}},\ \bibinfo {pages} {6119} (\bibinfo {year}
  {2009})}\BibitemShut {NoStop}%
\bibitem [{\citenamefont {Nemoto}\ \emph {et~al.}(2017)\citenamefont {Nemoto},
  \citenamefont {Guevara~Hidalgo},\ and\ \citenamefont
  {Lecomte}}]{Nemoto_2017}%
  \BibitemOpen
  \bibfield  {author} {\bibinfo {author} {\bibfnamefont {T.}~\bibnamefont
  {Nemoto}}, \bibinfo {author} {\bibfnamefont {E.}~\bibnamefont
  {Guevara~Hidalgo}}, \ and\ \bibinfo {author} {\bibfnamefont {V.}~\bibnamefont
  {Lecomte}},\ }\href {\doibase 10.1103/PhysRevE.95.012102} {\bibfield
  {journal} {\bibinfo  {journal} {Phys. Rev. E}\ }\textbf {\bibinfo {volume}
  {95}},\ \bibinfo {pages} {012102} (\bibinfo {year} {2017})}\BibitemShut
  {NoStop}%
\bibitem [{\citenamefont {Hidalgo}(2018)}]{Hidalgo_2018}%
  \BibitemOpen
  \bibfield  {author} {\bibinfo {author} {\bibfnamefont {E.~G.}\ \bibnamefont
  {Hidalgo}},\ }\href {\doibase 10.1088/1742-5468/aad6b2} {\bibfield  {journal}
  {\bibinfo  {journal} {Journal of Statistical Mechanics: Theory and
  Experiment}\ }\textbf {\bibinfo {volume} {2018}},\ \bibinfo {pages} {083211}
  (\bibinfo {year} {2018})}\BibitemShut {NoStop}%
\bibitem [{\citenamefont {Edelman}\ \emph {et~al.}(2016)\citenamefont
  {Edelman}, \citenamefont {Guionnet},\ and\ \citenamefont
  {P{\'e}ch{\'e}}}]{Edelman_2016}%
  \BibitemOpen
  \bibfield  {author} {\bibinfo {author} {\bibfnamefont {A.}~\bibnamefont
  {Edelman}}, \bibinfo {author} {\bibfnamefont {A.}~\bibnamefont {Guionnet}}, \
  and\ \bibinfo {author} {\bibfnamefont {S.}~\bibnamefont {P{\'e}ch{\'e}}},\
  }\href {\doibase 10.1214/15-AAP1129} {\bibfield  {journal} {\bibinfo
  {journal} {The Annals of Applied Probability}\ }\textbf {\bibinfo {volume}
  {26}},\ \bibinfo {pages} {1659 } (\bibinfo {year} {2016})}\BibitemShut
  {NoStop}%
\bibitem [{\citenamefont {Bolhuis}\ \emph {et~al.}(2002)\citenamefont
  {Bolhuis}, \citenamefont {Chandler}, \citenamefont {Dellago},\ and\
  \citenamefont {Geissler}}]{bolhuis2002}%
  \BibitemOpen
  \bibfield  {author} {\bibinfo {author} {\bibfnamefont {P.~G.}\ \bibnamefont
  {Bolhuis}}, \bibinfo {author} {\bibfnamefont {D.}~\bibnamefont {Chandler}},
  \bibinfo {author} {\bibfnamefont {C.}~\bibnamefont {Dellago}}, \ and\
  \bibinfo {author} {\bibfnamefont {P.~L.}\ \bibnamefont {Geissler}},\ }\href
  {\doibase 10.1146/annurev.physchem.53.082301.113146} {\bibfield  {journal}
  {\bibinfo  {journal} {Ann.\ Rev. Phys. Chem.}\ }\textbf {\bibinfo {volume}
  {53}},\ \bibinfo {pages} {291} (\bibinfo {year} {2002})},\ \bibinfo {note}
  {pMID: 11972010}\BibitemShut {NoStop}%
\bibitem [{\citenamefont {Hartmann}(2002)}]{align2002}%
  \BibitemOpen
  \bibfield  {author} {\bibinfo {author} {\bibfnamefont {A.~K.}\ \bibnamefont
  {Hartmann}},\ }\href {\doibase 10.1103/PhysRevE.65.056102} {\bibfield
  {journal} {\bibinfo  {journal} {Phys. Rev. E}\ }\textbf {\bibinfo {volume}
  {65}},\ \bibinfo {pages} {056102} (\bibinfo {year} {2002})}\BibitemShut
  {NoStop}%
\bibitem [{\citenamefont {Rohwer}\ \emph {et~al.}(2015)\citenamefont {Rohwer},
  \citenamefont {Angeletti},\ and\ \citenamefont {Touchette}}]{Rohwer_2015}%
  \BibitemOpen
  \bibfield  {author} {\bibinfo {author} {\bibfnamefont {C.~M.}\ \bibnamefont
  {Rohwer}}, \bibinfo {author} {\bibfnamefont {F.}~\bibnamefont {Angeletti}}, \
  and\ \bibinfo {author} {\bibfnamefont {H.}~\bibnamefont {Touchette}},\ }\href
  {\doibase 10.1103/PhysRevE.92.052104} {\bibfield  {journal} {\bibinfo
  {journal} {Phys. Rev. E}\ }\textbf {\bibinfo {volume} {92}},\ \bibinfo
  {pages} {052104} (\bibinfo {year} {2015})}\BibitemShut {NoStop}%
\bibitem [{\citenamefont {Bhamidi}\ \emph {et~al.}(2015)\citenamefont
  {Bhamidi}, \citenamefont {Hannig}, \citenamefont {Lee},\ and\ \citenamefont
  {Nolen}}]{Bhamidi_2015}%
  \BibitemOpen
  \bibfield  {author} {\bibinfo {author} {\bibfnamefont {S.}~\bibnamefont
  {Bhamidi}}, \bibinfo {author} {\bibfnamefont {J.}~\bibnamefont {Hannig}},
  \bibinfo {author} {\bibfnamefont {C.~Y.}\ \bibnamefont {Lee}}, \ and\
  \bibinfo {author} {\bibfnamefont {J.}~\bibnamefont {Nolen}},\ }\href
  {\doibase 10.1214/EJP.v20-2696} {\bibfield  {journal} {\bibinfo  {journal}
  {Electronic Journal of Probability}\ }\textbf {\bibinfo {volume} {20}},\
  \bibinfo {pages} {1 } (\bibinfo {year} {2015})}\BibitemShut {NoStop}%
\bibitem [{\citenamefont {Peliti}\ and\ \citenamefont
  {Pigolotti}(2021)}]{Peliti_2021}%
  \BibitemOpen
  \bibfield  {author} {\bibinfo {author} {\bibfnamefont {L.}~\bibnamefont
  {Peliti}}\ and\ \bibinfo {author} {\bibfnamefont {S.}~\bibnamefont
  {Pigolotti}},\ }\href {https://books.google.de/books?id=9rAPEAAAQBAJ} {\emph
  {\bibinfo {title} {Stochastic Thermodynamics: An Introduction}}}\ (\bibinfo
  {publisher} {Princeton University Press},\ \bibinfo {year}
  {2021})\BibitemShut {NoStop}%
\bibitem [{\citenamefont {O'Connell}(1998)}]{OConnell_1998}%
  \BibitemOpen
  \bibfield  {author} {\bibinfo {author} {\bibfnamefont {N.}~\bibnamefont
  {O'Connell}},\ }\href {\doibase 10.1007/s004400050149} {\bibfield  {journal}
  {\bibinfo  {journal} {Probability Theory and Related Fields}\ }\textbf
  {\bibinfo {volume} {110}},\ \bibinfo {pages} {277} (\bibinfo {year}
  {1998})}\BibitemShut {NoStop}%
\end{thebibliography}%
\end{document}